\documentclass[fleqn,usenatbib,useAMS]{mnras}

\usepackage{graphicx}	% Including figure files
\usepackage{amsmath}	% Advanced maths commands
\usepackage{amssymb}	% Extra maths symbols
\usepackage{multicol}        % Multi-column entries in tables
\usepackage{bm}		% Bold maths symbols, including upright Greek
\usepackage{pdflscape}	% Landscape pages
\usepackage[utf8]{inputenc}

\usepackage[dvipsnames]{xcolor}

\usepackage{soulutf8} % strikethrough com cor
\usepackage[T1]{fontenc}
\usepackage{ae,aecompl}

\usepackage{txfonts}
\title[Recovering EAs in CRTS]
{\center Recovering variable stars in large surveys:\\ EA$_{\rm up}$ Algol-type class in the Catalina Survey}

\author[Carmo et al.]
{A. Carmo$^{1}$\thanks{Contact e-mail: \href{aysses:ayssesdocarmo@gmail.com}{ayssesdocarmo@gmail.com}},
C. E. Ferreira Lopes$^{1}$,
A. Papageorgiou$^{2,3}$,
F. J. Jablonski$^{1}$,
\newauthor
C. V. Rodrigues$^{1}$,
A. J. Drake$^{4}$,
N. J. G.~Cross$^{5}$,
M. Catelan$^{2,3}$
\\
$^{1}$Instituto Nacional de Pesquisas Espaciais (INPE), Av. dos Astronautas, 1758 – São José dos Campos – SP, 12227-010, Brazil\\
$^{2}$Pontificia Universidad Católica de Chile, Facultad de Física, Instituto de Astrofísica Facultad de Física Av. Vicuña Mackenna 4860, 7820436 Macul,\\Santiago, Chile\\
$^{3}$Millennium Institute of Astrophysics, Santiago, Chile\\
$^{4}$California Institute of Technology, 1200 E. California Blvd,Pasadena, CA 91125, USA\\
$^{5}$ SUPA (Scottish Universities Physics Alliance) Wide-Field Astronomy Unit, Institute for Astronomy, School of Physics and Astronomy,\\ University of Edinburgh, Royal Observatory, Blackford Hill, Edinburgh EH9 3HJ, UK\\
}
\date{Accepted XXX. Received YYY; in original form ZZZ}
\date{Accepted 2020 August 13. Received 2020 July 26; in original form 2019 December 26}
\pubyear{2019}

\begin{document}
\label{firstpage}
\pagerange{\pageref{firstpage}--\pageref{lastpage}}
\maketitle

\begin{abstract}
The discovery and characterization of Algol eclipsing binaries (EAs) provide an opportunity to contribute for a better picture of the structure and evolution of low-mass stars. However, the cadence of most current photometric surveys hinders the detection of EAs since the separation between observations is usually larger than the eclipse(s) duration and hence few measurements are found at the eclipses. Even when those objects are detected as variable, their periods can be missed if an appropriate oversampling factor is not used in the search tools. In this paper, we apply this approach to find the periods of stars cataloged in the Catalina Real-Time Transient Survey (CRTS) as EAs having unknown period (EA$_{\rm up}$). As a result, the periods of $\sim 56\%$ of them were determined. Eight objects were identified as low-mass binary systems and modeled with the Wilson \& Devinney synthesis code combined with a Monte-Carlo Markov Chain optimization procedure. The computed masses and radii are in agreement with theoretical models and show no evidence of inflated radii. This paper is the first of a series aiming to identify suspected binary systems in large surveys.
% 189 words, normally of not more than 250 words for Main Journal papers
\end{abstract}

%one and six key 
%%%%%%%%%%%%%%%%%%%%%%%%%%%%%%%%%%%%%%%%%%%%%%%%%%

%%%%%%%%%%%%%%%%% BODY OF PAPER %%%%%%%%%%%%%%%%%%

% The MNRAS class isn't designed to include a table of contents, but for this document one is useful.
% I therefore have to do some kludging to make it work without masses of blank space.
\begingroup
\let\clearpage\relax
%\tableofcontents
\endgroup
\newpage
\section{Introduction}
\label{introduction}	

Eclipsing binary systems (EBs) give us important clues about the fundamental basis of stellar evolution since stellar quantities such as mass, radius, and temperature of the components can be directly assessed \citep{andersen1991,torres2010}. Until the end of the twentieth century, EBs were almost exclusively studied on a case-by-case basis \citep{mowlavi2017}. However, in recent years, the quality and quantity of astronomical data have significantly improved. Projects such as MACHO \citep{alcock1996}, OGLE \citep{udalski1992}, WFCAM \citep{hambly2008}, {\em Kepler} \citep{borucki2010}, CoRoT \citep{deleuil2018}, {\em Gaia} \citep{Gaiard2}, VVV \citep[][]{Minniti-2010}, TESS \citep{ricker2015}, NEOWISE \citep{NEOWISE}, and in the next few years, PLATO \citep[][]{Rauer-2014} and LSST \citep{abell2009,Ivezic-2019}, detected and will detect a large number of variable sources and a great effort is being made to provide tools for the analysis of such huge data sets.

EBs are grouped into three main branches according to the General Catalog of Variable Stars \citep[CGVS;][]{kazarovets2017general}: Algol (EA), $\beta$ Lyrae (EB) and W Ursae Majoris (EW). In particular, several astrophysical processes like interaction between components, mass transfer and magnetic braking can be investigated using EA systems \citep[e.g.][]{quian2018}. Another important feature of EA systems is that they may contain low-mass stars whose radii and masses can be known to better than 5\% accuracy \citep[often better than 3\%; see ][]{feiden2015}. There is a suggestion that radii of low-mass are inflated by more than 10\% in comparison with theoretical models for isolated stars \citep[][]{kraus2011,birkby2012,feiden2012,garrido2019}. This may be associated with non-solar metallicity \citep{berger2006,lopez-morales2007}, increased magnetic activity \citep{chabrier2007,kraus2011}, or be an observational effect related to distortions in the light curves caused by spots or flares \citep{morales2008,morales2010}. Only a small number of low-mass binaries that are detached EBs with components of late-K or M types present accurate radii determinations in the literature \citep{garrido2019}. 

Most of the known EAs are catalogued in databases such as the OGLE \citep{2016soszynski}; the GCVS \citep{kazarovets2017general}; the International Variable Star Index (VSX) by the American  Association  of  Variable  Star  Observers \citep[AAVSO;][]{watson2006}; the All-Sky Automated Survey \citep[ASAS;][]{richards_catalog2012}; the LAMOST survey \citep{quian2018}; the LINEAR survey \citep{palaversa2013}; and the Catalina Real-Time Transient Survey (CRTS) \citep{Drake-2014}. Although the number of catalogued EAs is fairly large, the observational strategies in the surveys vary a lot, and the methodologies for finding periods may fail in cases having narrow eclipses or few data points during these events. \citet{lopes2018} discussed the influence of the resolution of the frequency grid in the search for periodicities in EAs. They have succeeded in finding periods for 4 objects previously identified by \citet{Drake-2014} as having insufficient number of observations in the eclipses. 

This work is the first of a series in an attempt to determine the periods of EAs having narrow eclipses adopting the methodology proposed by \citet{lopes2018}. Here we focus on EAs from CRTS with unknown period (EA$_{\rm up}$ class). We have a particular interest in investigating the relatively rare low-mass binary systems. This paper is organized as follows. Section ~\ref{thedata} presents the sample of objects analyzed. Section ~\ref{search} presents a general discussion on the methods of periodic signals search. In section ~\ref{searchEAup}, three methods of period search are applied to the sample of interest. Section ~\ref{discussions} shows the characteristics of the EAs found in this work, the criteria to select low-mass stars, the method to obtain the individual stellar parameters and an analysis of low-mass stars in the radius inflation context. Finally, the conclusions are presented in Section ~\ref{conclusions}.
 
\section{Data: The CRTS EA sample}
\label{thedata}	
The Catalina Real-Time Transient Survey\footnote{\url{http://crts.caltech.edu/}} consists of a collaboration in which three telescopes are used aiming to discover Near-Earth Objects (NEOs) and Potential Hazardous Asteroids (PHAs)\footnote{\url{https://catalina.lpl.arizona.edu}}. The project covers the sky in the range of declinations $\delta=[-75^\circ,+65^\circ]$ and avoids crowded regions near to the Galactic plane ($|b| < 15^\circ$). The images are unfiltered to maximize the throughput and the photometry is carried out using the SExtractor photometry package \citep{bertin1996}. Based on data from the  0.7-m Catalina Schmidt Survey (CSS) telescope, \citet{Drake-2014} identified 47,000 variable sources from the public Catalina Data Release 1 \cite[CSDR1;][]{Drake-2012}, which together with objects from the other surveys produced an on-line catalog of 61,000 variable objects. Among them, 4680  objects are classified as EA binary systems. From the re-analysis of \citet{papageorgiou2018}, there are 3456 {\em bona fide} EA detached systems in that sample; they will be used for run time evaluations in a following subsection in this paper. For the scope of this work, we focus on the 153 EA$_{\rm up}$ systems listed in \citet{Drake-2014}, which were classified as unknown-period eclipsing binary candidates. In other words, these are objects that present variability typical of eclipsing variables (i.e., excursions to lower states of brightness) but had an insufficient number of observations in the eclipses for full characterization.

\section{Methodology for periodic signals search}
\label{search}

A first step in mining variable stars in large photometric surveys is the detection of changes in a source's brightness. Once the objects presenting variability have been found, a second step is the search for periodicities \citep[e.g.][]{Wozniak-2000,Shin-2009,FerreiraLopes-2015wfcam,FerreiraLopes-2016papI,FerreiraLopes-2017papII}.  

The identification of variability does not guarantee that the object is periodically variable. In this sense, EA-type stars can be missed both in target selection or in the periodicity search. The latter happens when the number of measurements at the eclipses is small and because outside the eclipses the variations are essentially due to noise. 

The periodicity search methods usually applied to astronomical time series rely on figures of merit. In terms of the associated phase diagram at each frequency grid point, these figures of merit measure correlations or some sort of ordering. Several works discuss the efficiency in finding periodic signals in astronomical time series \citep[e.g.][]{Heck1985,Swingler1989,Schwarzenberg1999,Shin2004}. \citet{Graham-2013} tested 11 different methods for 78 types of variable stars. They find that the phase dispersion-based techniques give the best results, but there are clear dependencies on object class and light-curve quality. 

In this work, three common period search methods were used: the Generalized Lomb-Scargle periodogram \citep[GLS;][]{Lomb-1976,Scargle-1982,Zechmeister-2009}, the String Length method \citep[STR;][]{Dworetsky-1983}, and the Phase Dispersion Minimization method \citep[PDM;][]{Stellingwerf-1978,Stellingwerf-2011}. The three methods mentioned above have figures of merit based on different premises. The performances of these methods to detect EA periodicities are compared in this section.

For unevenly spaced data, as is the case for most of the present surveys, we face an additional difficulty: the Nyquist frequency, $f_{Ny}$, has not a precise definition anymore, and the frequency grid is consequently not well defined as well. \citet{lopes2018} derived an expression (see Eq.~\ref{eq:lopes18}) that parameterizes the frequency resolution in terms of an oversampling factor with respect to the ideal, equally-spaced times series case. The number of frequencies $N_f$ is given by:

\begin{equation}
N_{f} = \frac{(f_{max}-f_{min}) \times T_{tot}}{\delta_{\phi}}, \label{eq:lopes18}
\end{equation}

\noindent where $f_{max}$ and $f_{min}$ are the maximum and minimum search frequencies, $T_{tot}$ is the total time baseline of the observations, and $\delta_{\phi}$ is a parameter that measures meaningful phase variations in the phase diagram when considering changes in frequency, $f$. We see that $\delta_\phi = 1$ corresponds to the minimum frequency sampling for an equally-spaced times series when $f_{min}=0$ and $f_{max}=f_{Ny}$. The quantity $1/\delta_{\phi}$ is called oversampling factor and has been used in expressions similar to Eq.~\ref{eq:lopes18} to define the frequency grid \citep{SchwarzenbergCzerny-1996,Debosscher2007,richards2012,VanderPlas-2015,VanderPlas-2018}. The $f_{min}$ value, even though being formally zero for equally spaced data, is usually defined as $2/T_{tot}$ to include at least two cycles of the longest period searched in the time series. $f_{max}$ is the upper limit in frequency, and for an equally spaced times series with step $\delta t$, $f_{max}=f_{Ny}=0.5/\delta t$. Prior knowledge on the shape of the periodic signal allows us to go far beyond the Nyquist limit, even for the ill-defined case of unevenly spaced data. Examples are the empirical values such as $f_{max}=10 \, {\rm d}^{-1}$ \citep{Debosscher2007, richards2012, DeMedeiros2013} or even larger \citep{Schwarzenberg1996,Damerdji2007,lopes2015a,FerreiraLopes2020}. 

%%%%%%%%%%%%%%%%%%%%%%%%%%%%%%%%%%%%%%%%%%%%%
\begin{figure}
\includegraphics[width=0.46\textwidth,height=0.48\textwidth,trim = 1mm 1mm 0mm 0mm,clip]{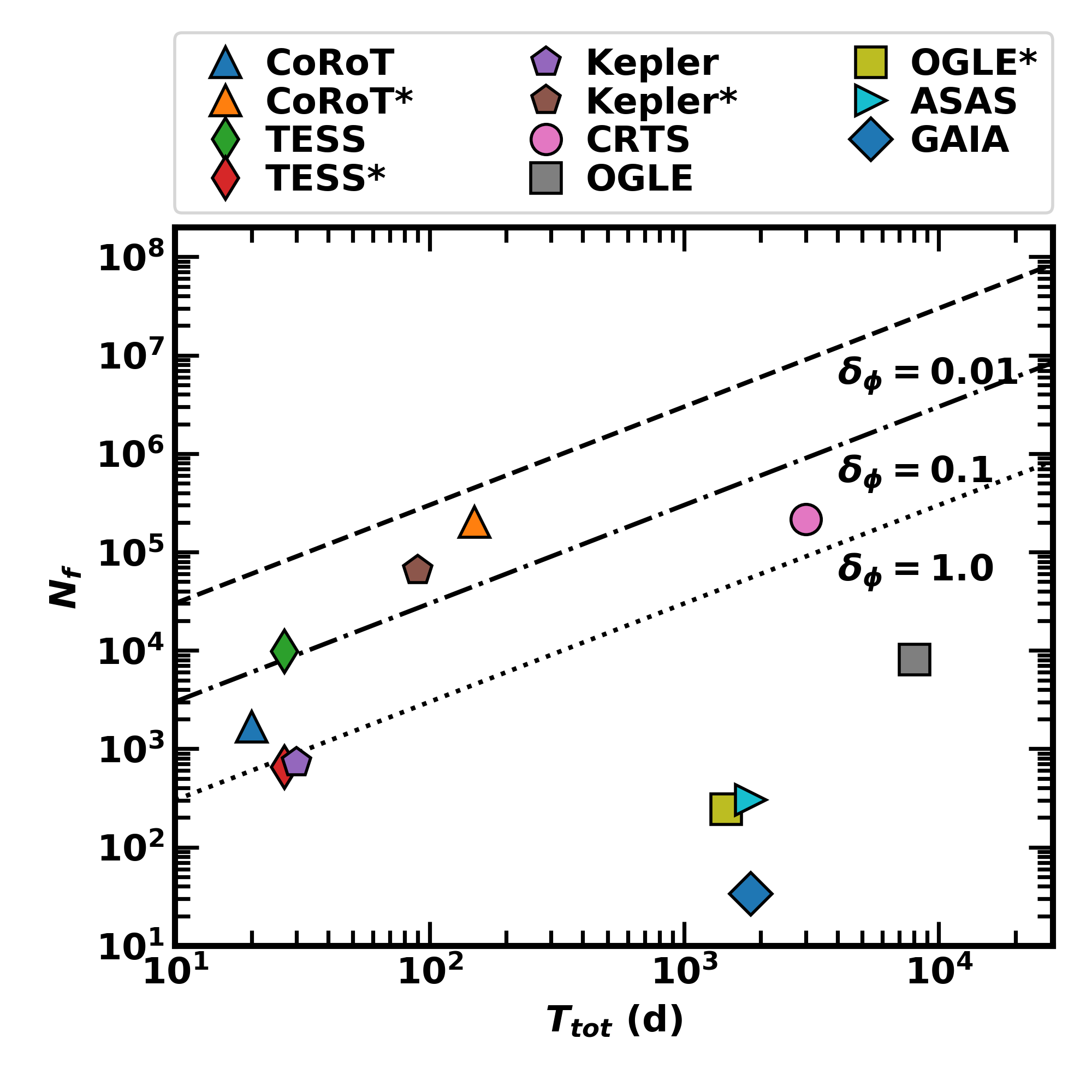}
\caption{The number of frequencies, $N_f$, versus total observation time, $T_{tot}$, for a number of recent surveys. The symbol {\em *} marks different combinations of cadences and total observing time. 
$N_f$ assumes a {\em median} interval between observations, $\delta t$, which would imply a Nyquist frequency $1/(2\delta t)$ for an equally spaced time series. The dotted line indicates a desirable sampling of $f_{max} = 30 {\rm \,d}^{-1}$, with $\delta_\phi$ from Eq.~\ref{eq:lopes18} set to 1. The vertical distance from the points to this line indicates how much the minimal grid of frequencies should be augmented to reach $30 {\rm \,d}^{-1}$ sampling. The dot-dashed and dashed lines correspond to oversampling factors of 10 and 100, or $\delta_\phi=0.1$ and $\delta_\phi=0.01$, respectively.}
\label{fig:freq}
\end{figure}
%%%%%%%%%%%%%%%%%%%%%%%%%%%%%%%%%%%%%%%%%%%%%%%%

The shape of the light curve has an important role in defining the grid of frequencies. Since eclipsing binaries in general have a strong first harmonic of the fundamental frequency, even in the ideal case of equally spaced data, the description of the light curve (e.g., in terms of Fourier components) would need at least twice the frequency sampling limit needed for the fundamental frequency alone. Figure 3 in \citet{lopes2018} illustrates how the choice of the oversampling factor ($1/\delta_{\phi}$) is important for detecting periodicities. It contains five phase diagrams of different types of variable stars. Detection of EAs is the most dependent on the oversampling factor. For instance, the use of a value of $\delta_{\phi}=0.2$, i.e., an oversampling of a factor of 5, still produces a blurred phase diagram.  This means that the frequency grid spacing should be finer to unambiguously identifying the correct orbital period of EAs.

Fig.~\ref{fig:freq} shows the plane $N_f$ (number of frequencies in the frequency grid) versus total observing time for a number of recent surveys, given the {\em median} sampling time, $\delta t$,  of each survey. The dotted line shows the locus of Eq.\,\ref{eq:lopes18} for a target $(f_{max}-f_{min})=30\, \rm{d}^{-1}$ and $\delta_\phi=1$. Values of the oversampling factor $1/\delta_\phi$ of 10 and 100 are shown for reference. Surveys such as {\em Kepler}, CoRoT and TESS have relatively small values of $T_{tot}$ but a good cadence, so they can even surpass the exemplified goal of having $(f_{max}-f_{min})=30\, \rm{d}^{-1}$. The other surveys have poorer cadences and require extending the natural frequency grid, besides oversampling it to probe time-scales of variability of the order of hours. One might ask what are the effects of integration time on the frequency domain. Since the measurements are mathematically equivalent to the convolution of Dirac-$\delta$ functions (the sampling) with a boxcar (the integration time), the result in the frequency domain is suppression of the high frequencies, as in the case for an equally spaced series. For the CRTS, the integration time is typically 30 sec and the median sampling is $\sim 20$ min, meaning that the suppression of high frequencies is small.

%%%%%%%%%%%%%%%%%%%%%%%%%%%%%%%%%%%%
\begin{figure*}
\includegraphics[width=0.45\textwidth,height=0.3\textwidth]{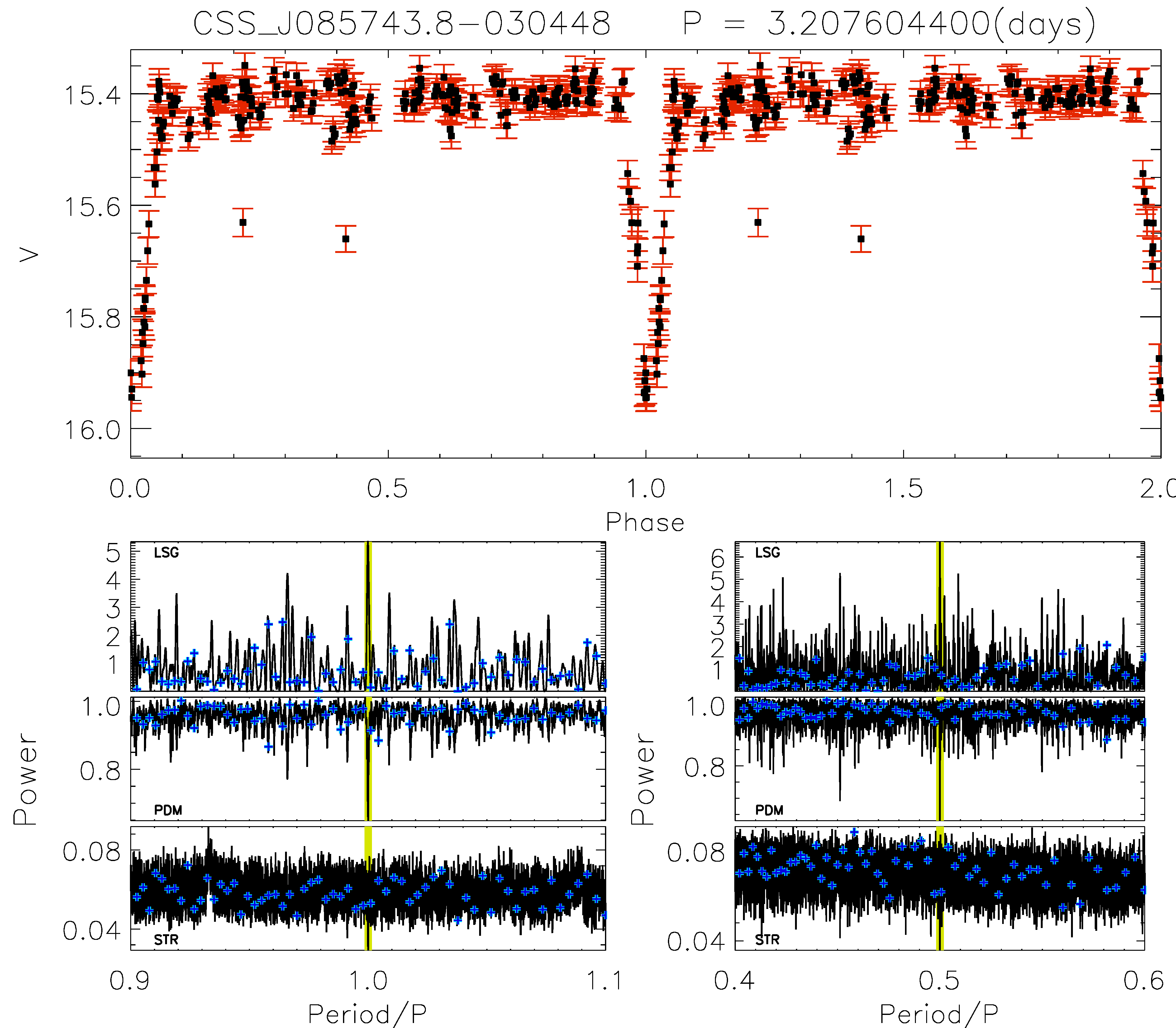}
\includegraphics[width=0.45\textwidth,height=0.3\textwidth]{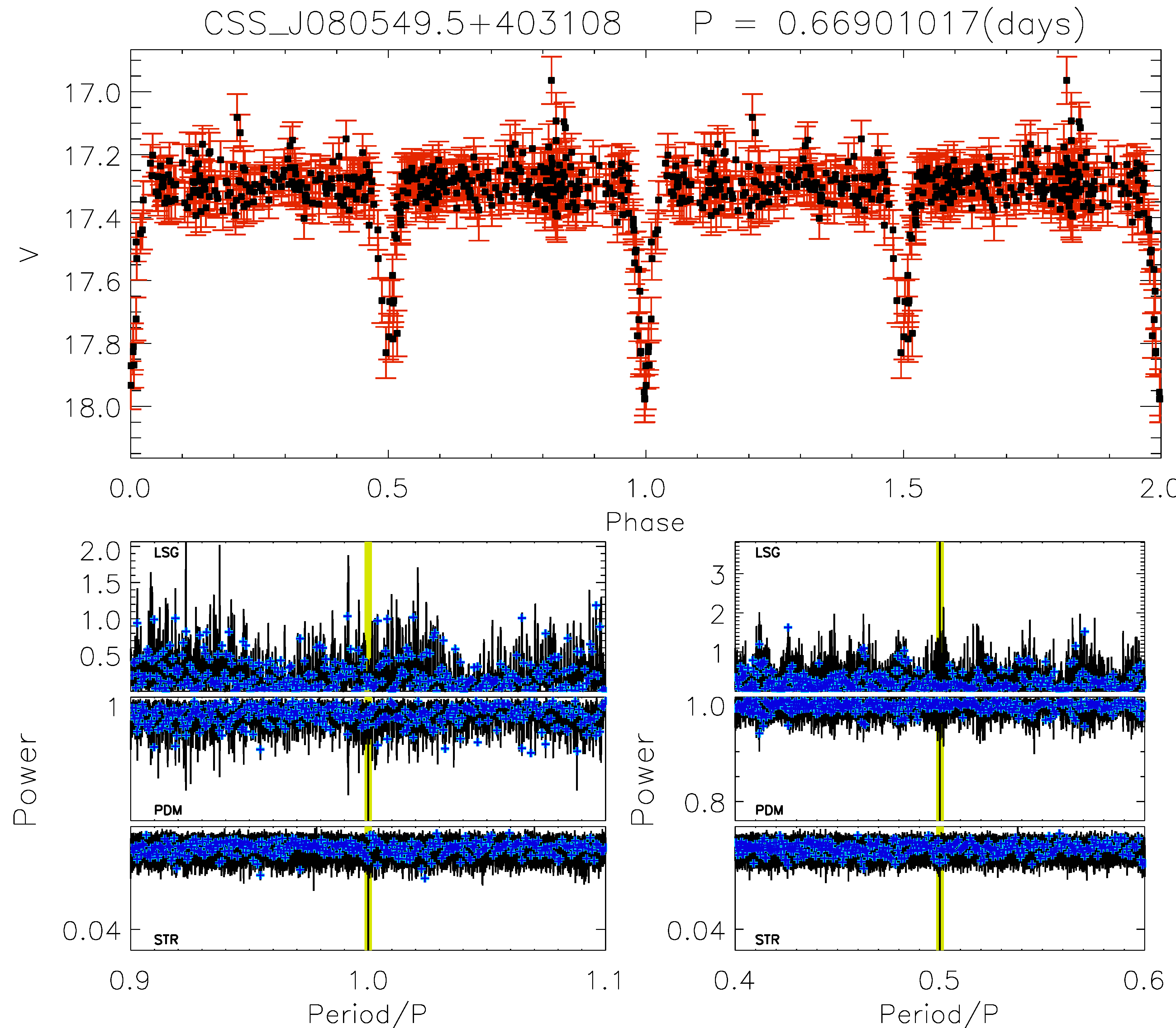}\\
\includegraphics[width=0.45\textwidth,height=0.3\textwidth]{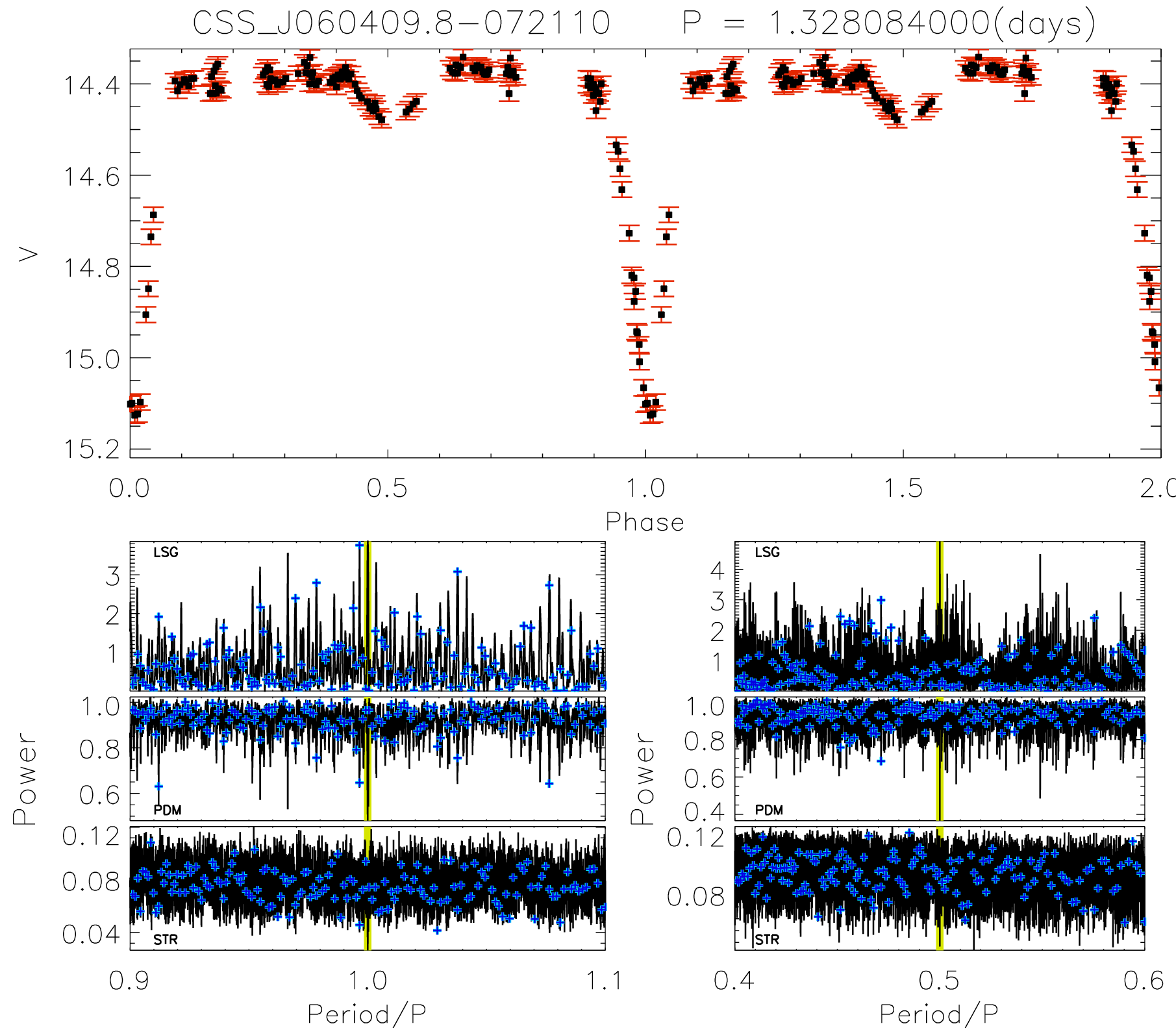}
\includegraphics[width=0.45\textwidth,height=0.3\textwidth]{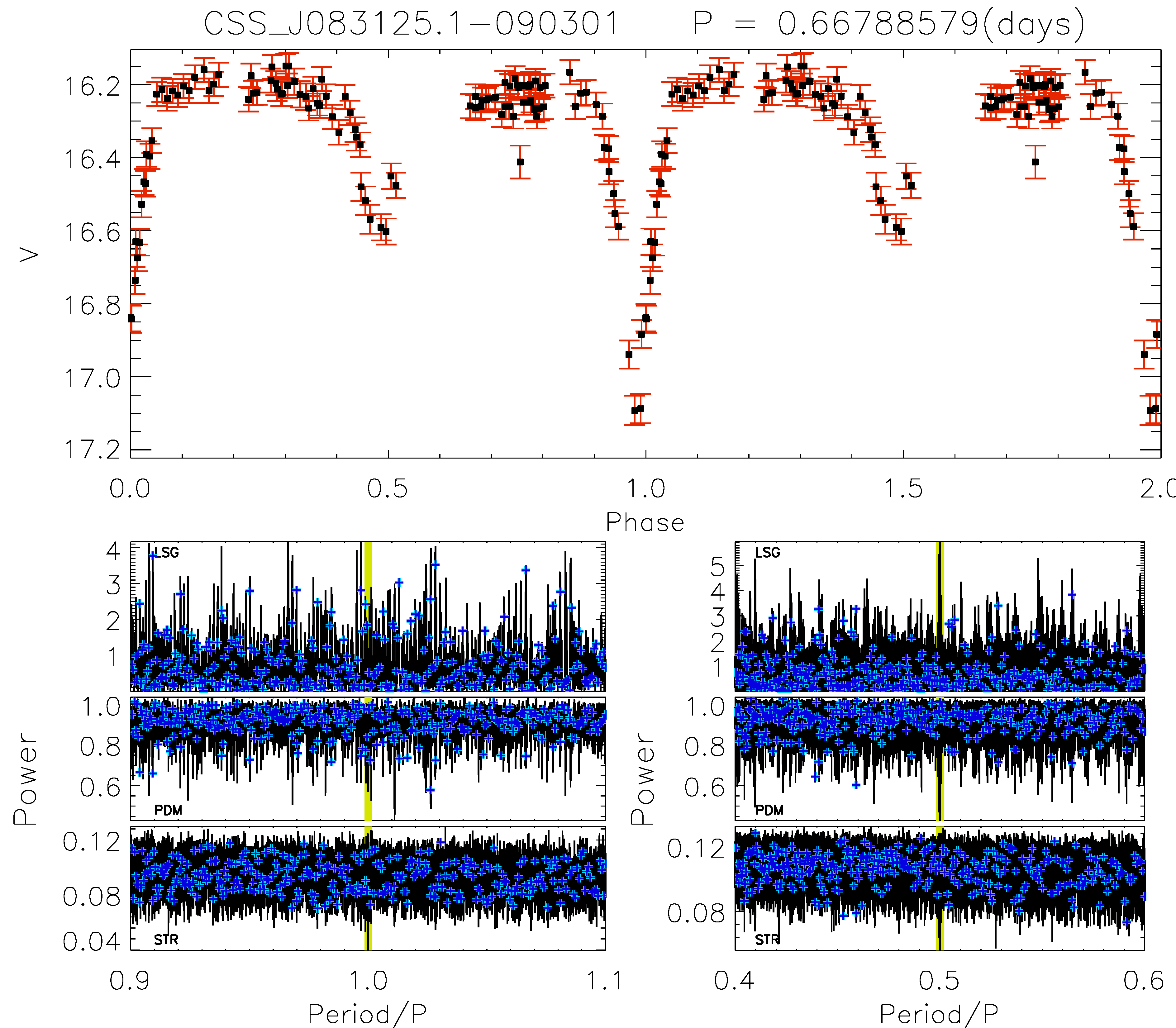}
\caption{Phase diagrams and periodograms for four EA$_{\rm up}$ CRTS objects with new period determinations. For each object, the bottom panels show the GLS, PDM, and STR periodograms for the main variability period $P$ (left panels) and the first harmonic, $0.5\times P$ (right panels). A vertical yellow line highlights the main variability period. The periodograms in black use the frequency grid adopted in this work ($\delta_\phi=0.01$). The blue crosses show the periodograms using a coarser grid. The object name and period found are shown at the top of the each diagram.}
\label{fig:periodograms}
\end{figure*}
%%%%%%%%%%%%%%%%%%%%%%%%%%%%%%%%%%%%

%%%%%%%%%%%%%%%%%%%%%%%%%%%%%%%%%%%%
\begin{table}
\caption[]{$E_{rate}$ for the GLS, PDM, and STR methods. Hits consist of fractional differences in the period identified with respect to the cataloged value of less than 1\%. The sample used consists of 3456 detached EAs cataloged by \citet{papageorgiou2018}.}
\label{perhit}
\centering
 \begin{tabular}{l c c c}
\hline \hline
Method & Hits (\%) & Multiples (\%) & Misses (\%)\\
\hline
GLS & 2 & 84 & 14\\
PDM & 11 & 82 & 7\\
STR & 50 & 33 & 17\\
\hline
\end{tabular}
\end{table}
%%%%%%%%%%%%%%%%%%%%%%%%%%%%%%%%%%%%%

\subsection{Run time and hit rate}
\label{sec_runtime}

To test with different values of $\delta_{\phi}$, we called the GLS, PDM, and STR procedures in a single loop for which the total run time of an object is the sum of the times spent to run each of the three methods. Obviously, the run time is directly proportional to the product $N_f \times Np$, where $Np$ is the number of points in the light curve. A fiducial mark for this is a run time of 13 sec to explore a dataset with $Np=312$ points, and $N_f=2 \times 10^5$ frequencies. We also define an efficiency rate, $E_{rate}$, which is calculated as follows.

We used the sample of 3456 detached EAs cataloged by \citet{papageorgiou2018} as a reference since those systems have well determined periods. We ran experiments with an oversampling factor of 100 (or $\delta_{\phi}$=0.01), and $N_f = 2 \times 10^6$. $E_{rate}$ is the fractional number of systems for which we can recover the correct orbital period with a difference less than 1\% from the cataloged value. Table \ref{perhit} summarizes the results for the different methods. The second column represents direct detections of the fundamental frequency and the third column corresponds to harmonics and sub-harmonics.

Periodicity search methods often find values that are half or multiples of the correct period when we take into account that the shape of the light curve of an eclipsing binary (specially detached systems) is very well defined. Cases of exactly equal components are relatively rare compared to the spurious cases. Hits at $1/2$ times the correct period are expected, since Fourier-based methods are sensitive to a single harmonic frequency, and the first harmonic of the fundamental orbital frequency may have a larger amplitude than the fundamental. This problem is known and discussed in the literature \citep[e.g.,][]{richards2012, Drake-2014}. Even the better behaved methods as STR and PDM may be fooled by the presence of particular configurations in the folded light curve or by chance arrangements due to noise. We see that the most reliable method in terms of direct hits is STR, which found the cataloged periodicity for 50\% of the objects in the test sample. If we add to this the detections of multiples, $E_{rate}$=83\%. The count of successes in this form is even better for PDM, with $E_{rate}=93\%$. GLS also performs well, with $E_{rate}= 85\%$. The three methods combined allow recovering 98\% of the periods.

The most common reasons for misses are strong levels of noise, outliers that deviate substantially from the mean noise characteristics and cases with period close to multiples of one day. The GLS method has only $\sim 2\%$ of hits, but this is not surprising considering that we are treating highly non-sinusoidal light curves with poor sampling in time. The PDM method has a better performance compared to GLS, finding the correct periodicity for 11\% of the objects in the test sample, but we have to recall that this method suffers in the case of small number of data points and big gaps in the phase diagram.

%%%%%%%%%%%%%%%%%%%%%%%%%%%%%%%
\begin{figure*}
\includegraphics[width=0.33\textwidth,height=0.3\textwidth]{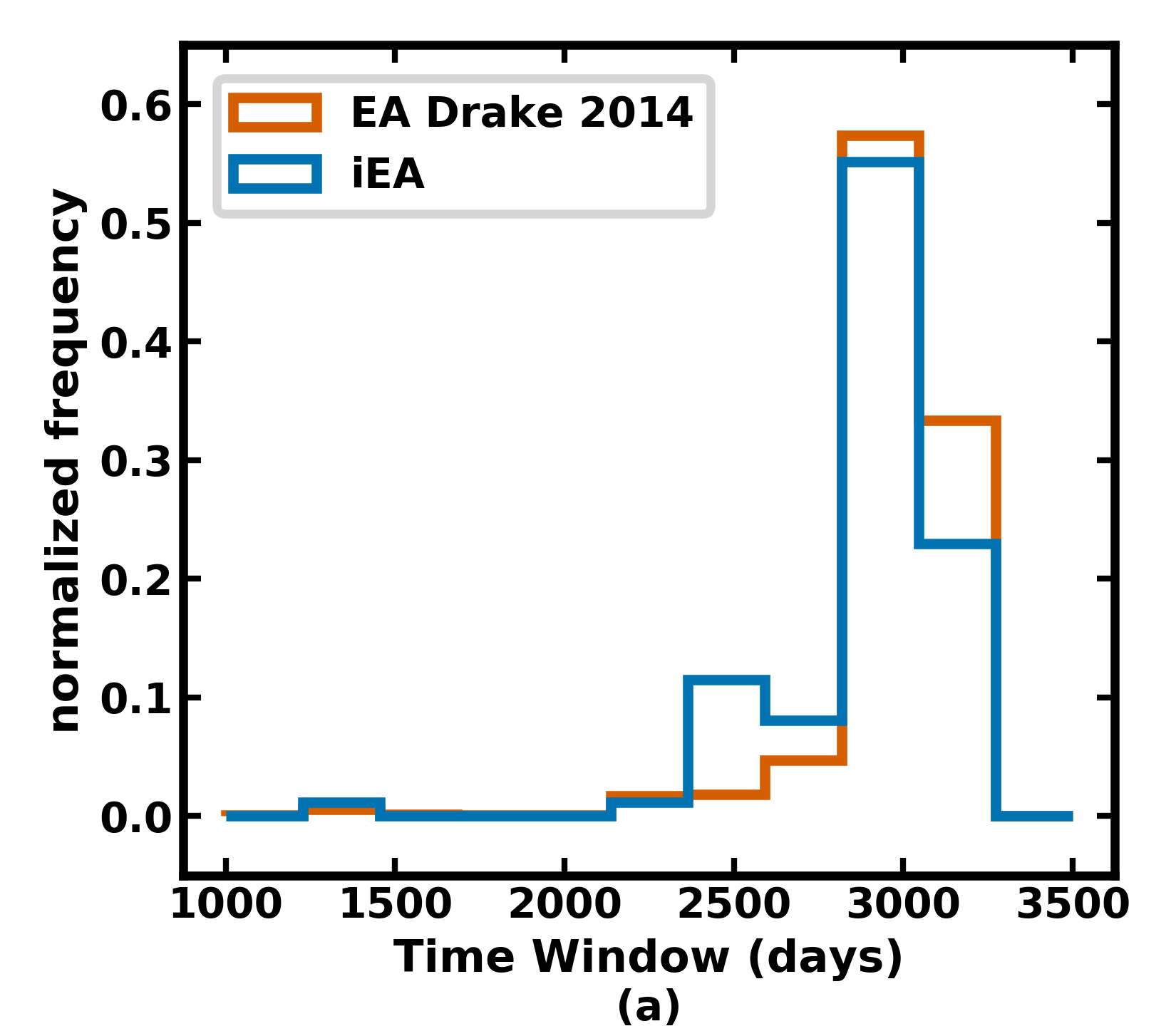}
\includegraphics[width=0.33\textwidth,height=0.3\textwidth]{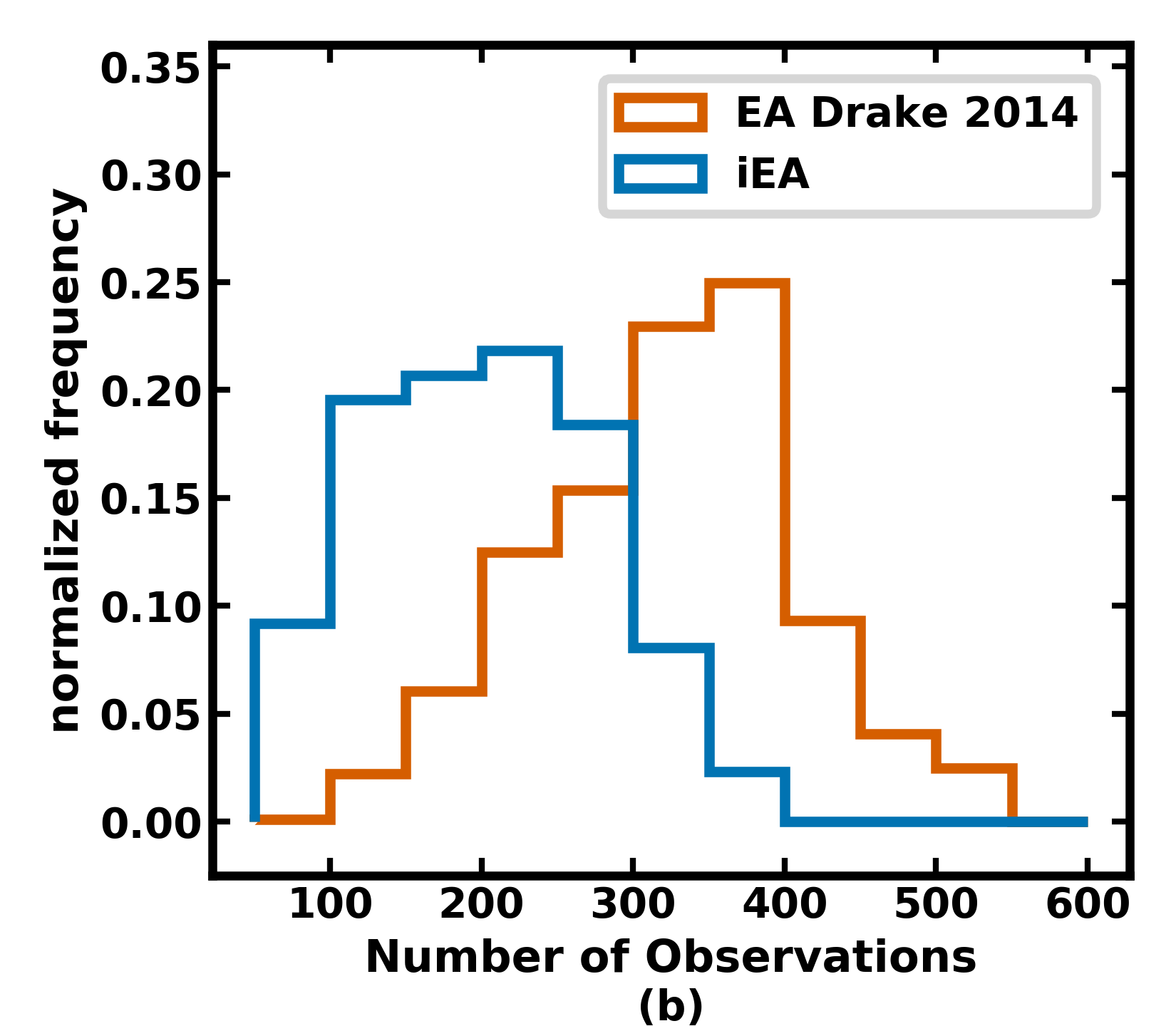} 
\includegraphics[width=0.33\textwidth,height=0.3\textwidth]{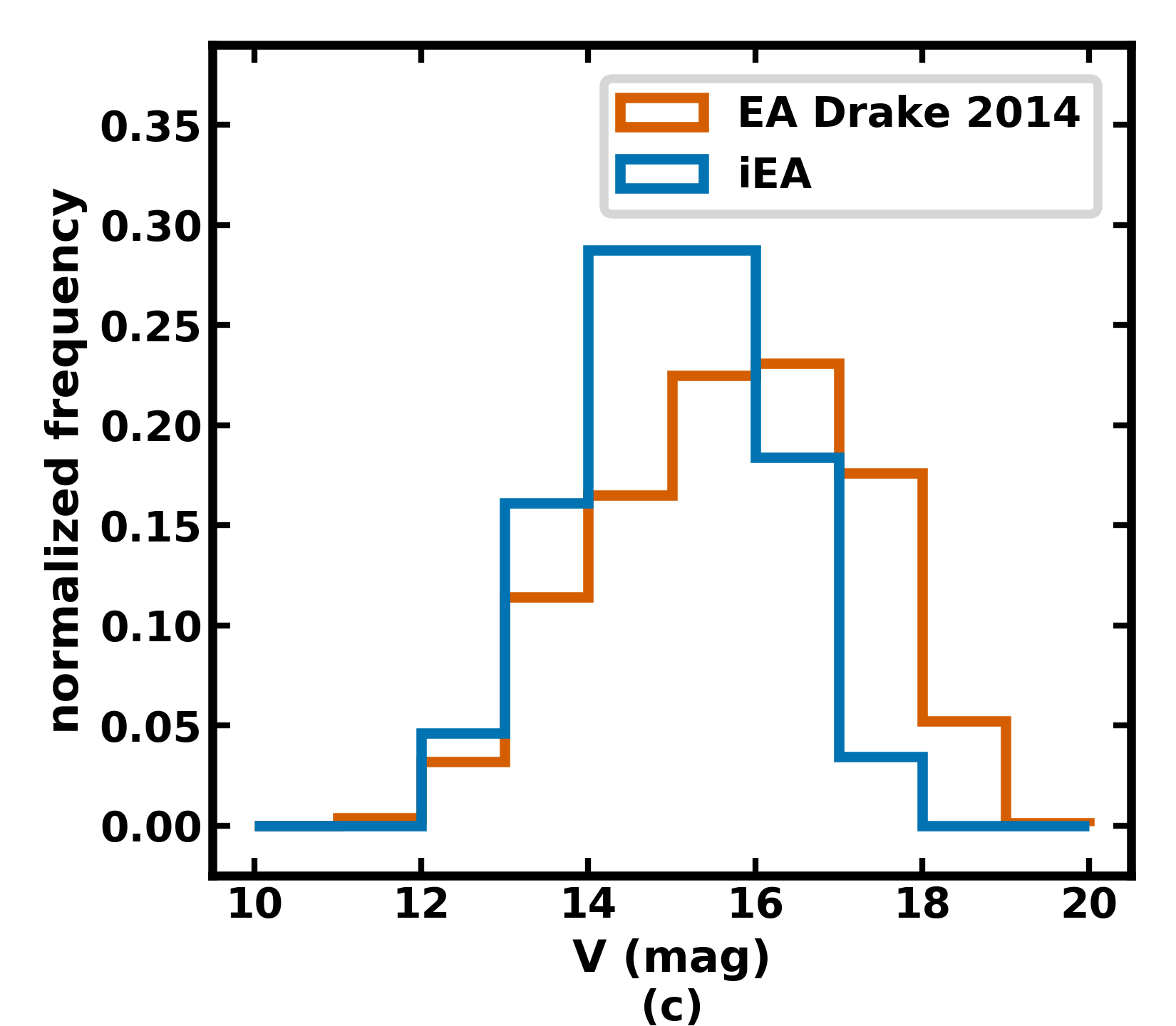}\\
\includegraphics[width=0.33\textwidth,height=0.3\textwidth]{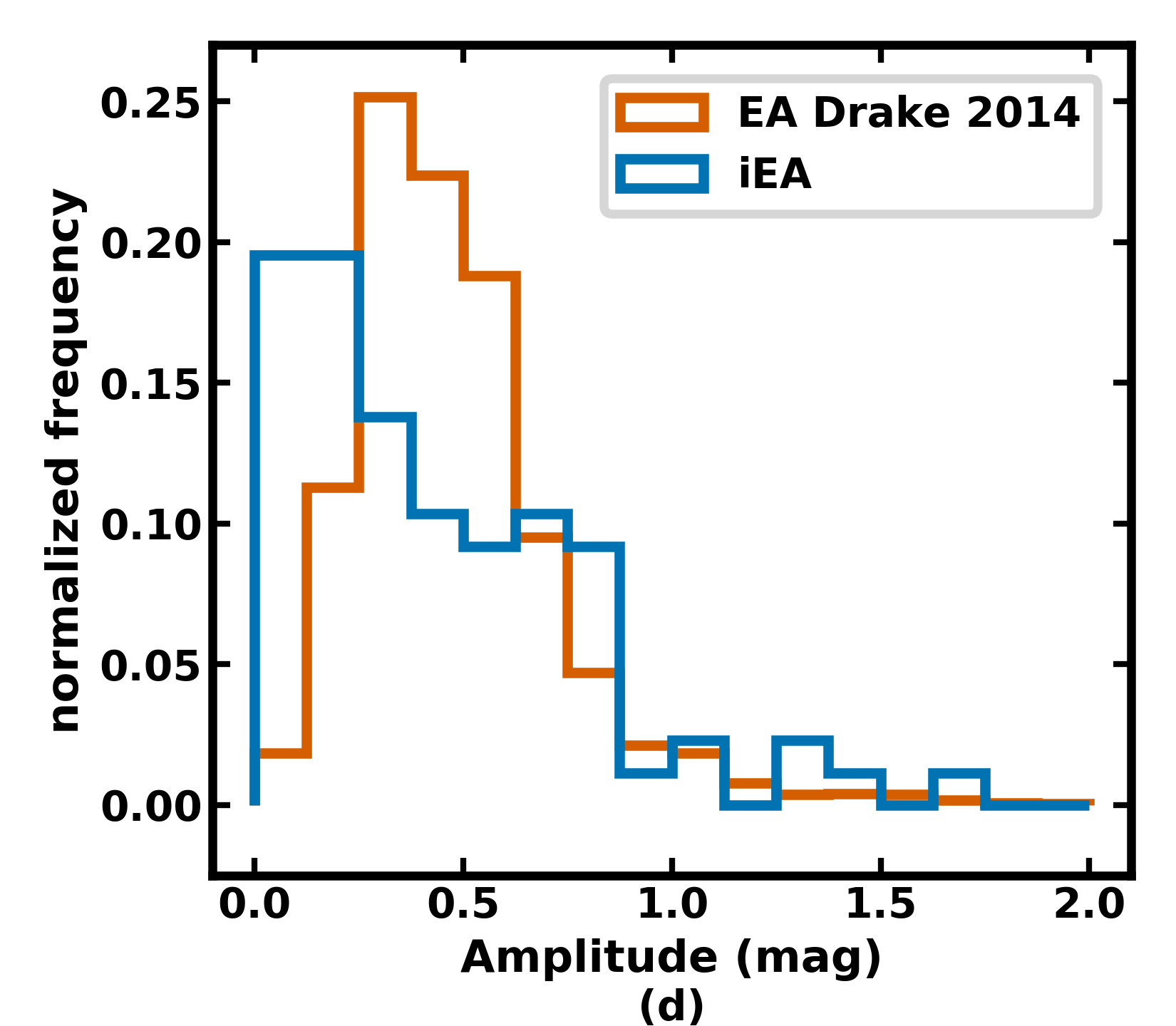}
\includegraphics[width=0.33\textwidth,height=0.3\textwidth]{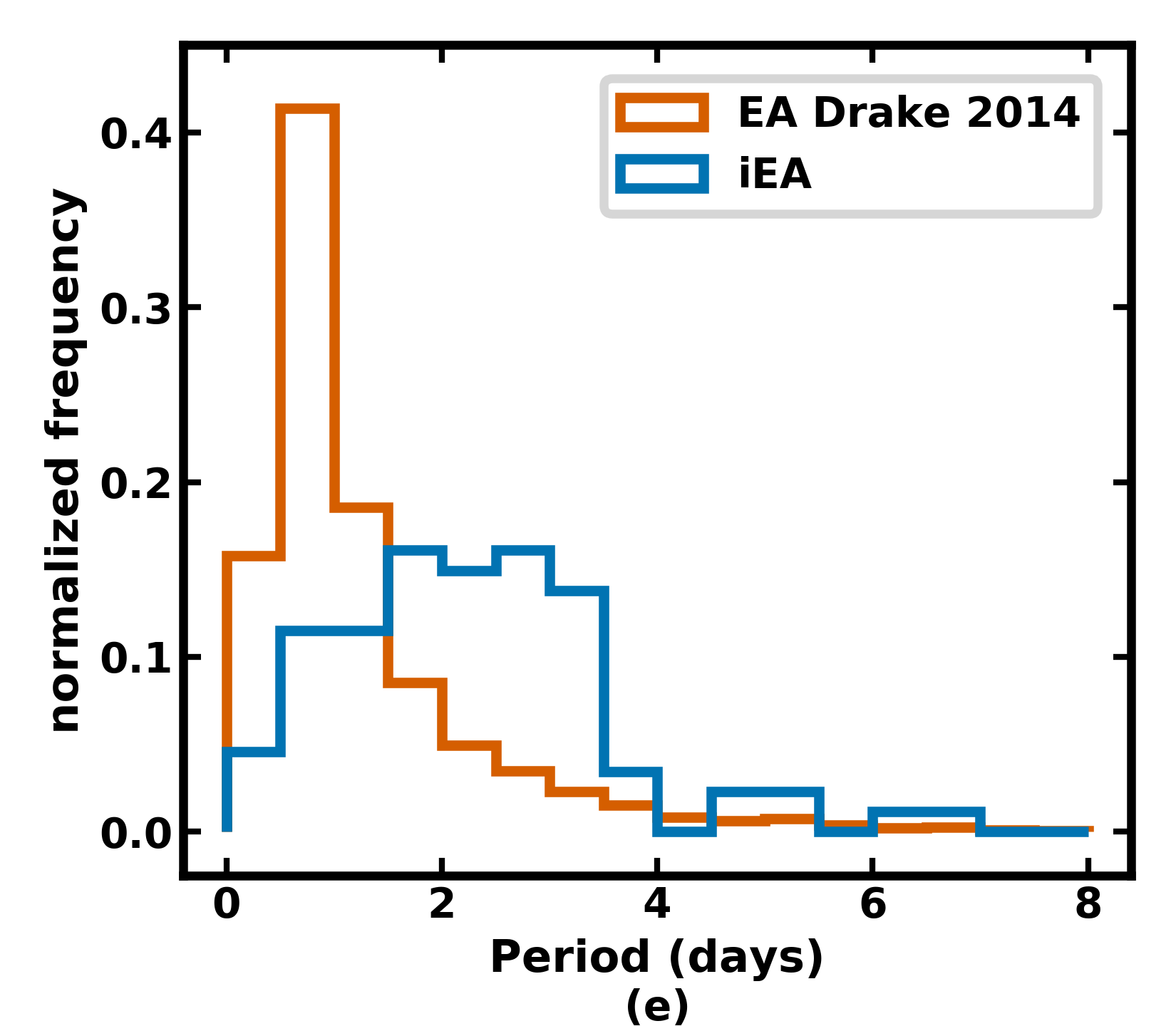}
\includegraphics[width=0.33\textwidth,height=0.3\textwidth]{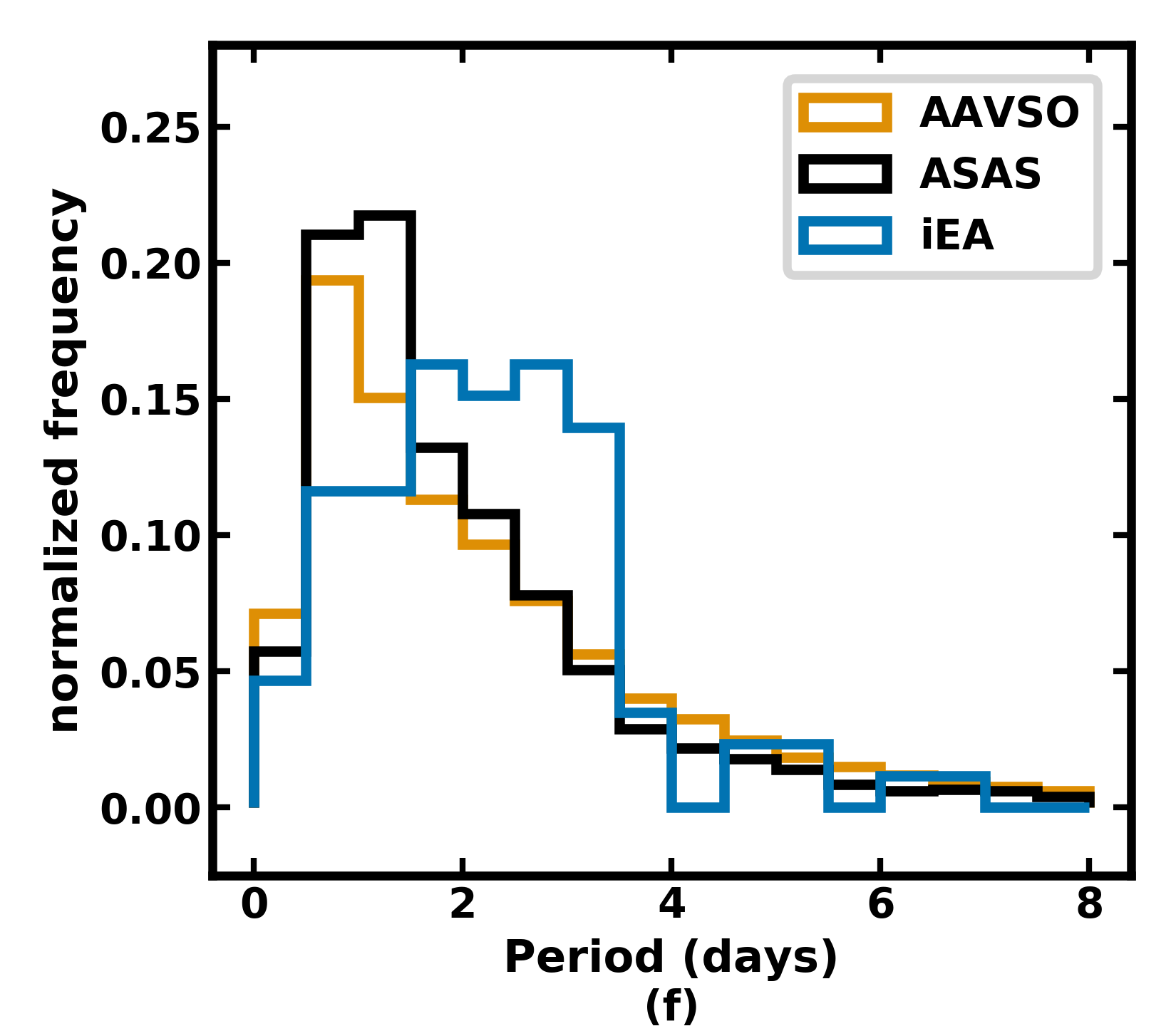}
\caption{Panels (a) to (e) compare the iEA sample with the catalog of EAs with known periods from \citet{Drake-2014}. Panel (f) compares the period distribution of the iEA with objects having periods up to 8~days in other catalogs from the literature (AAVSO and ASAS).}
\label{fig:histo_pers}
\end{figure*}
%%%%%%%%%%%%%%%%%%%%%%%%%%%%%%%

\section{Searching for the periods of EA$_{\rm up}$ stars}
\label{searchEAup}	

Guided by the benefits of oversampling and by careful examination of the phase diagrams once a signal is detected, we examined the folded light curves of the 153 EA$_{\rm up}$ listed in \citet{Drake-2014}.  Table~\ref{tab:periodsiEA} in the Appendix \ref{sec:tables1} shows the parameters related to the period search in the EA$_{\rm up}$ sample. Convincing light curves were obtained for 87 sources, i.e., 56\% of the EA$_{\rm up}$ sample. From now on, we will refer to this set of objects as "iEA": a shortcut for "sample of newly identified eclipsing binaries among the EA$_{\rm up}$ of \citet{Drake-2014}".

Figure~\ref{fig:periodograms} shows periodograms and folded light curves for four iEA objects. The lower panels illustrate how a fine frequency grid (in black, $\delta_\phi=0.01$) improves the detection of signals since the frequency grid points are closer to the true frequency. An inadequate, coarse frequency grid (in blue) may miss completely the correct peaks.

\section{Results and Discussion}
\label{discussions}

To have a broad view about the characteristics of the sample of newly identified objects, we compare in Fig. \ref{fig:histo_pers} the properties of iEA systems with those of the EAs previously identified by \citet{Drake-2014}. The main results for each panel are summarized below:

\begin{itemize}
    \item[(a)] The time windows of the iEA and EA samples did not present any significant difference, as expected;
    \item[(b)] The iEAs had a smaller number of observations in comparison with the EA sample, the medians being 200 and 332 observations, respectively. In other words, the iEAs had 40\% less observations than the EA sample; 
    \item[(c)] The median magnitudes were 15.08 mag and 15.85 mag for the iEA and EA samples, respectively. It means that the iEAs were slightly brighter than the EAs, on average by $0.77$ mag;
    \item[(d)] Our approach tends to be more efficient in EAs with depth of eclipse shallower than 0.25 mag;
    \item[(e)] The period distributions of iEAs and EAs showed median values of 2.12 days and 0.87 days, respectively. 
    This indicates we were identifying a number of long-period systems, which have a poor sampling due to the small number of points in the light curve; 
    \item[(f)] The same behavior found in panel (e) was seen when other catalogs are considered. The median periods within the range of 0-8 days were 1.49 days and 1.50 days, for the AAVSO respectively \citep{watson2006} and ASAS \citep{richards_catalog2012} catalogs.    
\end{itemize}

Overall, our approach detected EA systems having longer periods (lower number of cycles in the total time span of the observations) and smaller number of observations. We used the Mann-Whitney U-test statistics to evaluate if the iEA and EA samples belong to the same parent population. For the relevant parameters (panels c-f in Fig. \ref{fig:histo_pers}), the null hypothesis that the samples belong to the same parent population is not accepted at the 99\% confidence level. Given the sizes of the samples, this suggests that the entanglement of poor sampling in terms of cadence and small number of samples may have an important impact on any attempt to derive statistical properties from time-sparse surveys.

\subsection{Low-mass eclipsing binaries}
\label{findlowmass}

Given the importance of low-mass stars to study theoretical models of stellar structure and evolution, we used color criteria to look for low-mass stars within the iEA sample. The first step was to transform $V_{CSS}$ to Johnson $V$. We used the expression presented in \citet{drake2013}:
\begin{equation}
V=V_{CSS} + 0.31 \times (B-V)^2+0.04.
\label{eq:vcss}
\end{equation}
The $(B-V)$ index came from the APASS catalog \citep{Henden2016}. The photometric uncertainties of the original CSS data were determined using an empirical relationship between source flux and the observed photometric scatter. \citet{graham2017} presented a correction for the estimated error (see their Fig. 1). An analytical fit to this correction is shown in \citet{papageorgiou2018}, which we used here.

We adopted the color criteria of \citet{papageorgiou2018} to select low-mass stars candidates. They were: $V-K_{S}>3.0$, according to \citet{hartman2011}; $0.35<J-H<0.8$ and $H-K_{S} \leq 0.45$ mag, based on \citet{Lepine2011} and \citet{Zhong2015}. Infrared colors were obtained from the 2MASS $JHK_S$ photometry \citep{Cutri2003}. For both 2MASS and APASS catalogs, we performed a conservative search within a radius of 2\arcsec\ at the position of each iEA object. Twenty-two objects do not have APASS $(B-V)$ color information; for these, we apply the transformation from 2MASS colors to the Johnson-Cousins system, as provided by \citet[][their Eq. 16]{Bilir2008}.

%%%%%%%%%%%%%%%%%%%%%  COLOR INDEX
\begin{figure}
\includegraphics[width=0.522\textwidth,height=0.40\textwidth,trim = 4mm 0mm 8mm 8mm,clip]{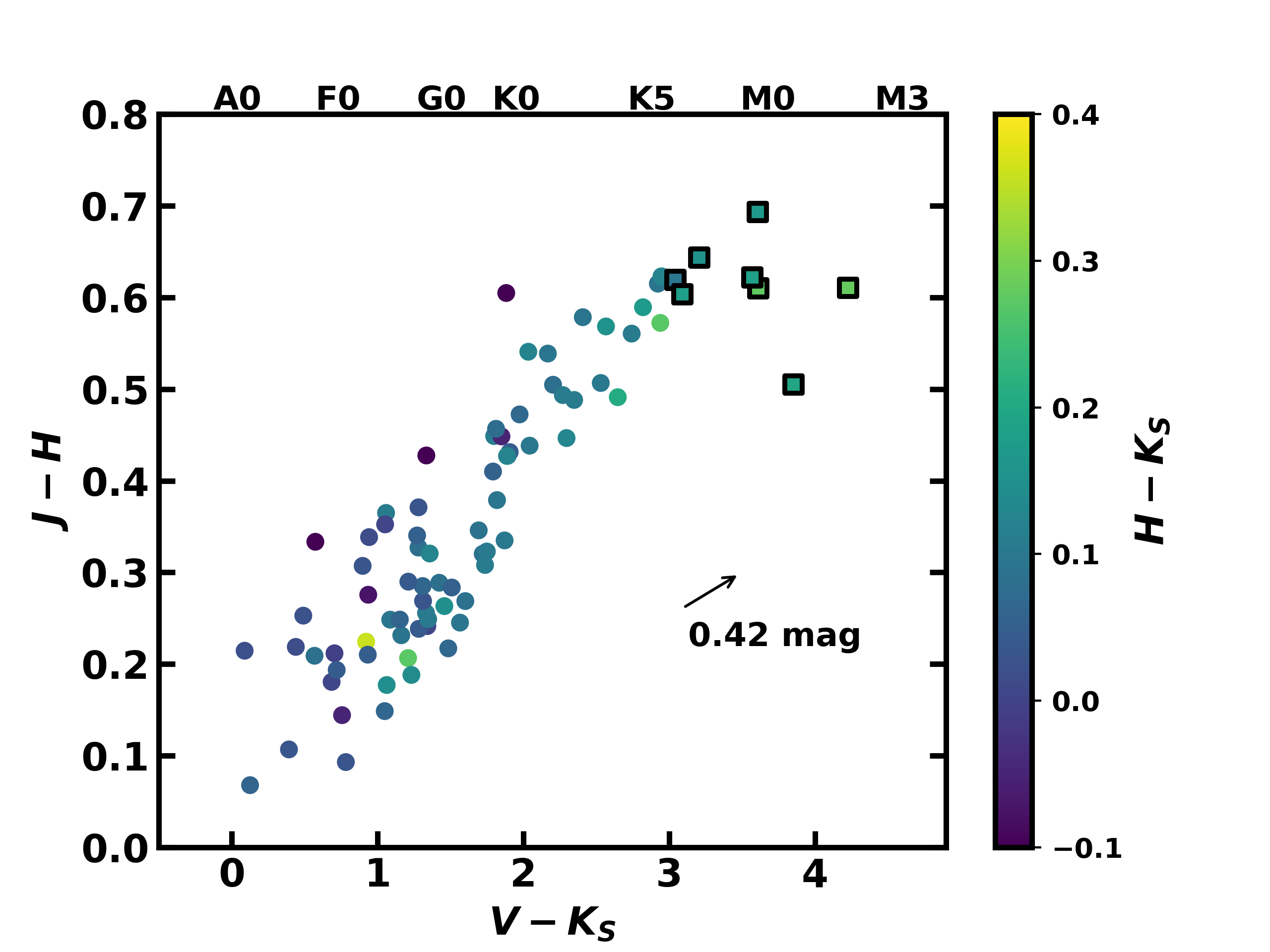}
\caption{Color-color diagram in the near-infrared for 87 iEA objects. The corresponding spectral type for a given $V-K_S$ according to \citet{pecaut2013} is shown in the top axis. The squares represent our eight low-mass candidates, which correspond to K5-M3 main-sequence stars. The arrow indicates the mean value of the extinction for the iEA sample.}
\label{fig:color-infrared}
\end{figure}
%%%%%%%%%%%%%%%%%%%%%

The literature data were collected from Vizier catalogs using the \textit{astroquery} modules that handle Vizier and CDS catalogs and Astropy tools\footnote{\url{http://www.astropy.org}} \citep {astropy:2013, astropy:2018}. The reddening value in the $V$ band was calculated using $E(B-V)$ values by \citet{green2018} and a total-to-selective extinction ratio $R_V=3.1$. The Python package \textit{dustmaps}\footnote{\url{https://doi.org/10.21105/joss.00695}} \citep{green2015} allows us to use the best-fit (maximum probability density) line-of-sight reddening corresponding to each distance modulus derived from the {\em Gaia} DR2. For the $JHK_S$ bands, the interstellar extinction was calculated using the Python package \textit{mwdust}\footnote{\url{https://github.com/jobovy/mwdust}} \citep{bovy2016}, which supports the combination of the following catalogs: \citet{marshall2006}, \citet{green2015} and \citet{drimmel2003}. Fig.~\ref{fig:color-infrared} shows the result of the color criteria applied for iEAs as a color-color diagram, in which 8 low mass system candidates are located.

\begin{figure}
\includegraphics[width=0.50\textwidth,height=0.40\textwidth,trim = 4mm 0mm 8mm 8mm,clip]{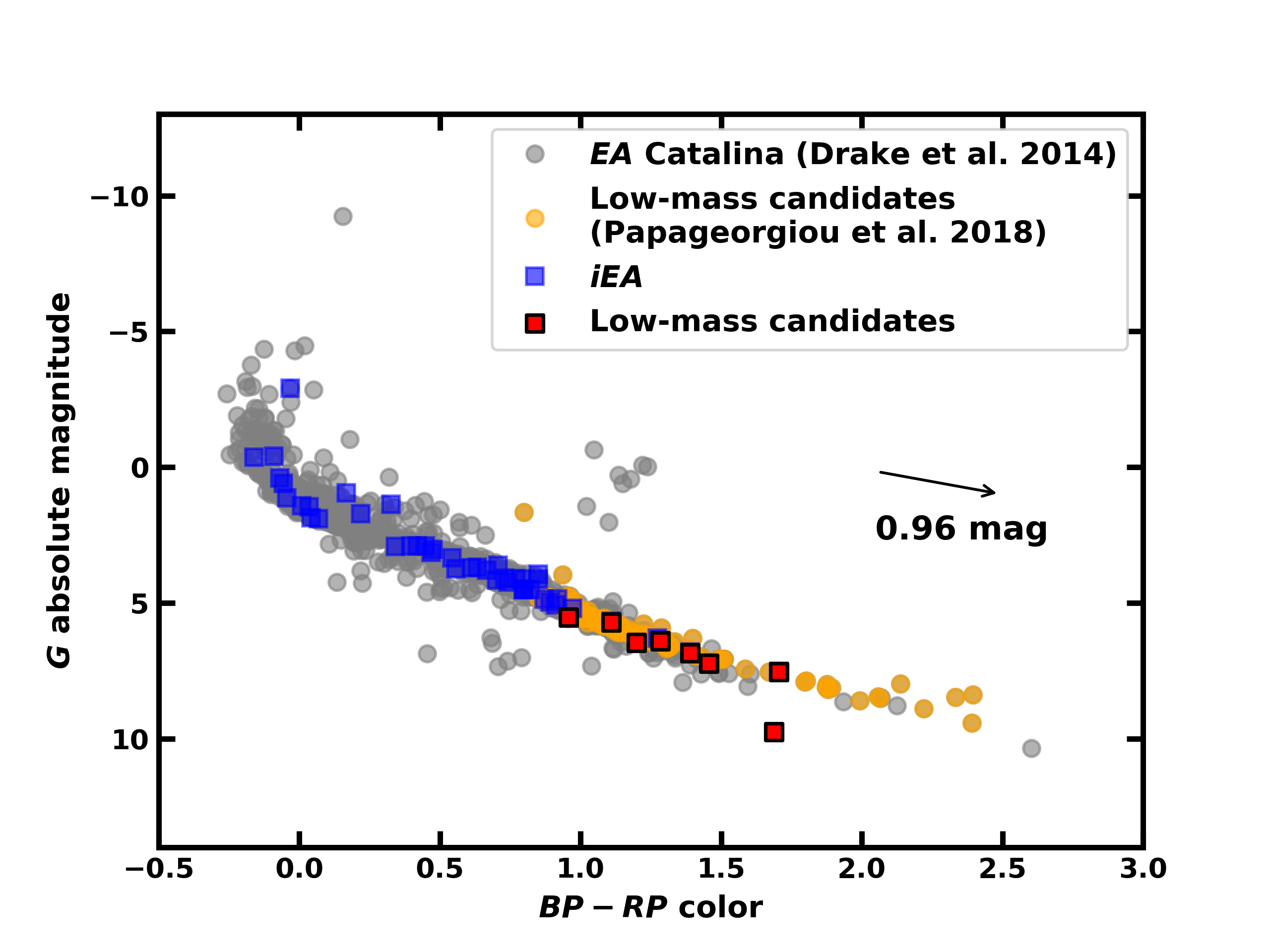}
\caption{Color-absolute magnitude diagram for different samples of EAs. The objects from the \citet{Drake-2014} catalog are depicted in gray dots. In yellow dots, objects selected by \citet{papageorgiou2018} as low-mass objects. The red squares are the low-mass iEA candidates from this work. The blue squares represent the remaining iEAs. The reddening vector was calculated from the mean value of the extinction for all iEAs.}
\label{fig:color-gaia}
\end{figure}

We used the Gaia data to build the color-absolute magnitude diagram of EAs as shown in Fig. \ref{fig:color-gaia}.
The extinction values $A_{G}$ and $E(G_{BP}-G_{RP})$ are obtained directly from Gaia DR2. 
For objects that do not have these values in the catalog, we use a correction based on values of $A_{V}=3.1 E(B-V)$, as given by
\citet[][see Eq. 1]{2018gaia}.
As we can see, the 8 iEAs low-mass candidates share the same region in the diagram as the low-mass systems reported by \citet{papageorgiou2018}. They have K-M spectral types with temperatures of about 4,000 K.

\subsection{Low-mass stellar parameters}
\label{parameters}

The light curves used in this step present long-term variations which were removed using a linear, parabolic, or sine function as in \citet[][see section 5]{papageorgiou2018}. Such variations can be related to stellar magnetic activity or to the presence of a third body in the system  \citep[e.g.][]{Applegate-1992,Morales-2010,FerreiraLopes-2015cycles,Bours-2016,Almeida-2019}. Besides that, the presence of a non-visible third body in an eccentric orbit can cause a rapid orbital precession \citep{soderhjelm1975}. Therefore, in observations with many cycles, the eclipses may become shallower and shallower a long time and even disappear \citep{graczyk2011,jurysek2018}. For the 8 low-mass stars found, objects CSS J084835.7 + 253917, CSS J020021.5 + 213412, and CSS J071357.2 + 342138 showed eclipses only in the first half part of the time series. Examining in detail, we see that the absence of eclipses in the second half of the light curves is consistent with the combination of the cadence of the observations and the orbital period itself.

The light curves of the low-mass iEA candidates were modeled with the \citet[WD;][]{wilson1971} code, which is widely used for the analysis of eclipsing binary data. The Markov Chain Monte Carlo \citep[MCMC;][]{Metropolis1953} method was used to find the modal values of the fitted parameters of the light curve synthesis code from the probability distribution of the parameters, and consequently, an estimate of the errors associated to each parameter. Parameter convergence happens in up to $\sim50,000$ iterations. The model had four free parameters: secondary temperature ($T_2$), modified Kopal potentials ($\Omega_1$, $\Omega_2$), and orbital inclination ($i$). The primary temperature $T_1$ was assumed to correspond to the $T_{\rm eff}$ value given in {\em Gaia} DR2. Only CSS J080549.6+403108 does not have an estimated temperature. In this case, we set $T_1$ based on the {\em Gaia} absolute magnitude (combined for the two stars) and the depth of the eclipses, as explained below. The mass ratio $q$ and eccentricity $e$ were fixed such that $q=1$ and $e=0$.

%%%%%%%%%%%%%%%%%%%  COLOR ABSOLUTE {\em Gaia}
%%%%%%%%%%%%%%%%%%%

To limit the range of values for the parameters in a model fit, we used the total magnitude of the system as an additional constraint. The predicted total apparent magnitude outside the eclipses is estimated taking into account the interstellar extinction and the distance. As in the previous section, we adopted the distances from \citet{bailer-jones2018} with the revisions from \citet{Gaiard2} and reddening from \cite{green2018}.  Here the morphological parameters of the phase-folded light curves do not need a physical modeling and they were obtained with the procedure LMfit-py \citep{newville2016}\footnote{\url{http://doi.org/10.5281/zenodo.11813}}, which provides a least-squares minimization routine for analysis of the data. Two Gaussians for the primary and secondary eclipses plus a constant out-of-eclipses baseline were used as the model. 

We assumed that the ratio of the depths of eclipses is proportional to the ratio of the individual luminosities: $\Delta I_1/\Delta I_2 \propto L_1/L_2$. Using this information, we estimated the spectral type of each component. We made use of the temperature-dependent bolometric corrections for stars in the main sequence of \citet{pecaut2013}. This provides a starting point to find a solution, and helps to define the range in which each parameter is searched for. The depth ratio is also used by other authors to estimate temperatures in eclipsing systems \citep[e.g.,][]{armstrong2014}. 

Regarding other model parameters, they were fixed according to the following assumptions. The albedos were assumed to be $A_1=A_2=0.5$, the gravity darkening coefficients were adopted as $g_1=g_2=0.32$ considering stars with convective envelopes \citep{lucy1967}, and the stellar limb-darkenings were set to a linear law. The third light parameter was set to $l_3=0$. 
The WD code was run in mode 2, suitable for detached systems.

%%%%%%%%%%%%%%%%%%%%
\begin{table*}
 \scriptsize
\caption{Absolute parameters of the stellar components of the low-mass binary system candidates. The errors were estimated considering an accuracy of the effective temperature from {\em Gaia} of 324~K.}
 \label{tab:abspars}
 \begin{tabular}{l c c c c c c c c c c}
\hline
Name & ID & Period (days) & $T_{\rm eff}$ (K) & $T_1$ (K) & $T_2$ (K) & $M_1$ ($M_\odot$) & $M_2$ ($M_\odot$) & $R_1$ ($R_\odot$) & $R_2$ ($R_\odot$) & $a$ ($R_\odot$)\\[0.5pt] 
\hline
CSS J003441.8-135033 & 1012004004843 & 1.97420921 & 3896 & 3938 & 3844 & $0.61^{+0.16}_{-0.16}$ & $0.58^{+0.09}_{-0.16}$ & $0.56^{+0.03}_{-0.05}$ & $0.53^{+0.03}_{-0.06}$ & $7.03^{+0.35}_{-0.69}$\\[3pt]
CSS J145100.7+052843 & 1104080068168 & 5.06655534 & 3954 & 3974 & 3932 & $0.62^{+0.13}_{-0.13}$ & $0.61^{+0.09}_{-0.13}$ & $0.43^{+0.02}_{-0.03}$ & $0.42^{+0.02}_{-0.03}$ & $13.33^{+0.62}_{-1.03}$\\[3pt]
CSS J162549.4+102124 & 1109087063294 & 2.07014986 & 3828 & 3853 & 3800 & $0.59^{+0.20}_{-0.20}$ & $0.57^{+0.10}_{-0.20}$ & $0.51^{+0.03}_{-0.06}$ & $0.49^{+0.03}_{-0.07}$ & $7.17^{+0.38}_{-0.93}$\\[3pt]
CSS J020021.5+213412 & 1121011041164 & 2.32385716 & 3819 & 3819 & 3819 & $0.58^{+0.20}_{-0.20}$ & $0.58^{+0.10}_{-0.20}$ & $0.57^{+0.03}_{-0.08}$ & $0.57^{+0.03}_{-0.08}$ & $7.74^{+0.41}_{-1.05}$\\[3pt]
CSS J084835.7+253917 & 1126043006161 & 2.43505230 & 4510 & 4587 & 4406 & $0.78^{+0.08}_{-0.08}$ & $0.74^{+0.08}_{-0.08}$ & $0.64^{+0.02}_{-0.02}$ & $0.60^{+0.02}_{-0.02}$ & $8.76^{+0.31}_{-0.32}$\\[3pt]
CSS J071357.2+342138 & 1135032018057 & 4.87375954 & 3868 & 3896 & 3835 & $0.60^{+0.17}_{-0.17}$ & $0.58^{+0.09}_{-0.18}$ & $0.62^{+0.03}_{-0.06}$ & $0.59^{+0.03}_{-0.07}$ & $12.79^{+0.65}_{-1.40}$\\[3pt]
CSS J080549.6+403108 & 1140034027271 & 0.66901017 & 3512 & 3514 & 3510 & $0.39^{+0.23}_{-0.23}$ & $0.39^{+0.19}_{-0.23}$ & $0.30^{+0.04}_{-0.08}$ & $0.30^{+0.04}_{-0.08}$ & $2.96^{+0.42}_{-0.76}$\\[3pt]
CSS J090355.4+533132 & 1152031059450 & 0.66327468 & 3962 & 3963 & 3961 & $0.62^{+0.13}_{-0.13}$ & $0.62^{+0.09}_{-0.13}$ & $0.61^{+0.03}_{-0.05}$ & $0.61^{+0.03}_{-0.05}$ & $3.44^{+0.16}_{-0.26}$\\[3pt]
\hline
\end{tabular}
\end{table*}
%%%%%%%%%%%%%%%%%%%%%%%%

The parameter ratios obtained from the WD code fitting of the light curves are very reliable. However, there is a large degeneracy in the determination of the absolute values of each component. To circumvent this problem, we followed the procedure described in \citet[][Sec. 5]{coughlin2011}. It combines the observed effective temperature of the entire system, parameter ratios from the WD fitting, and theoretical values of radii and temperatures of low-mass stars. We adopted the effective temperature from {\em Gaia}, which has a typical accuracy of 324~K, for sources brighter than $G = 17$ mag and having $T_{\rm eff}$ in the range 3,000 -- 10,000~K. We used this error for all eight low-mass candidates, including ${\rm CSS J}003441.8-135033$ and ${\rm CSS J}080549.6+403108$, which have $G \sim 17.3$~mag. From the WD fitting, we obtained the temperature ratio and $r_{sum}$, which is defined by \citet{coughlin2011} as the sum of the stellar radii divided by the binary's semi-major axis. The theoretical values of radius and temperature are those presented in \citet{baraffe1998}. We assumed an age of 5.0~Gyr and ${\rm [M/H]} = 0.0$ for $0.075 \leq M(M_{\odot}) \leq 1.0$.  The resulting absolute parameters are shown in Table~\ref{tab:abspars}.

\begin{figure*}
\includegraphics[width=0.33\textwidth,height=0.3\textwidth,trim = 1mm 1mm 1mm 0mm,clip]{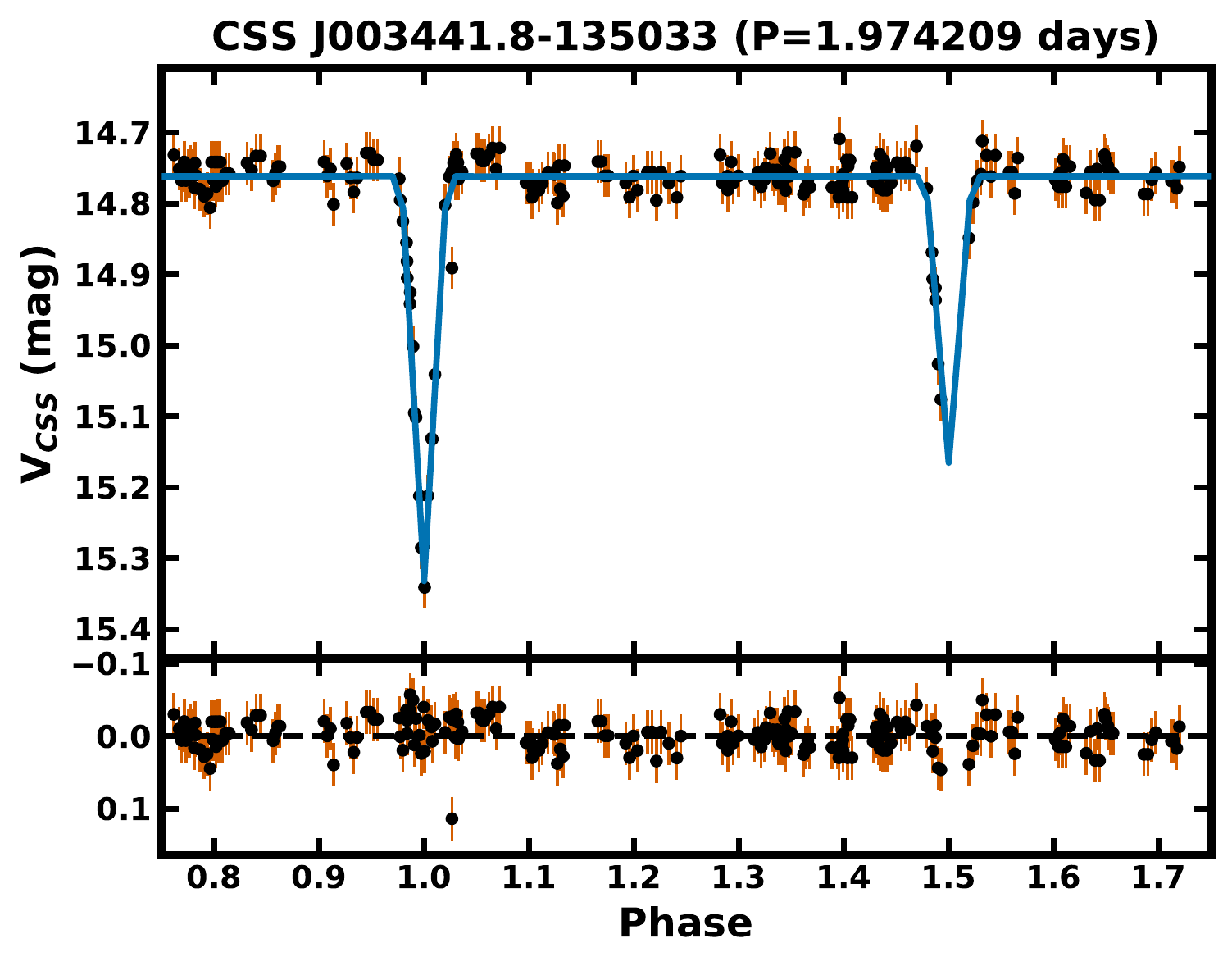}
\includegraphics[width=0.33\textwidth,height=0.3\textwidth,trim = 1mm 1mm 1mm 0mm,clip]{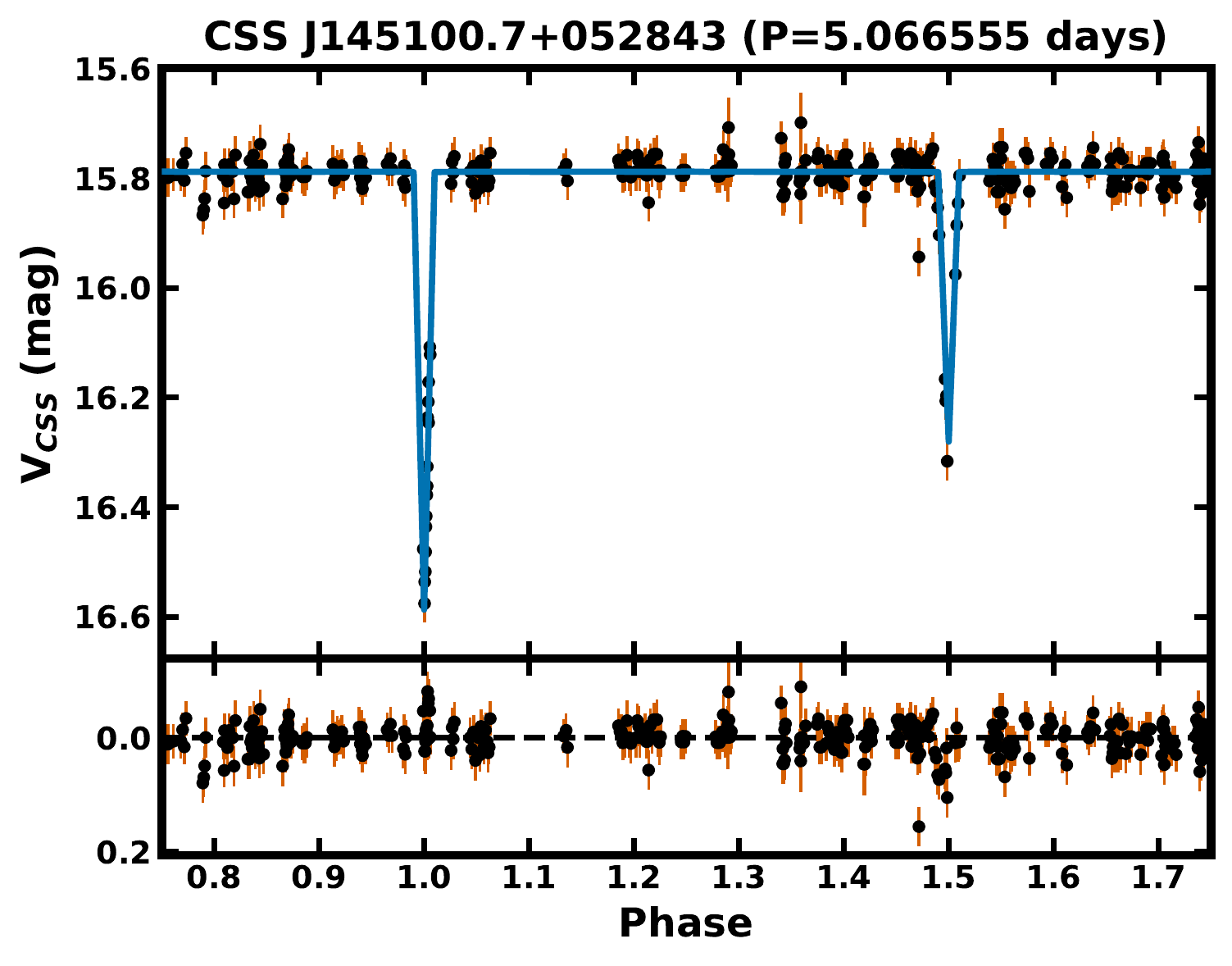}
\includegraphics[width=0.33\textwidth,height=0.3\textwidth,trim = 1mm 1mm 1mm 0mm,clip]{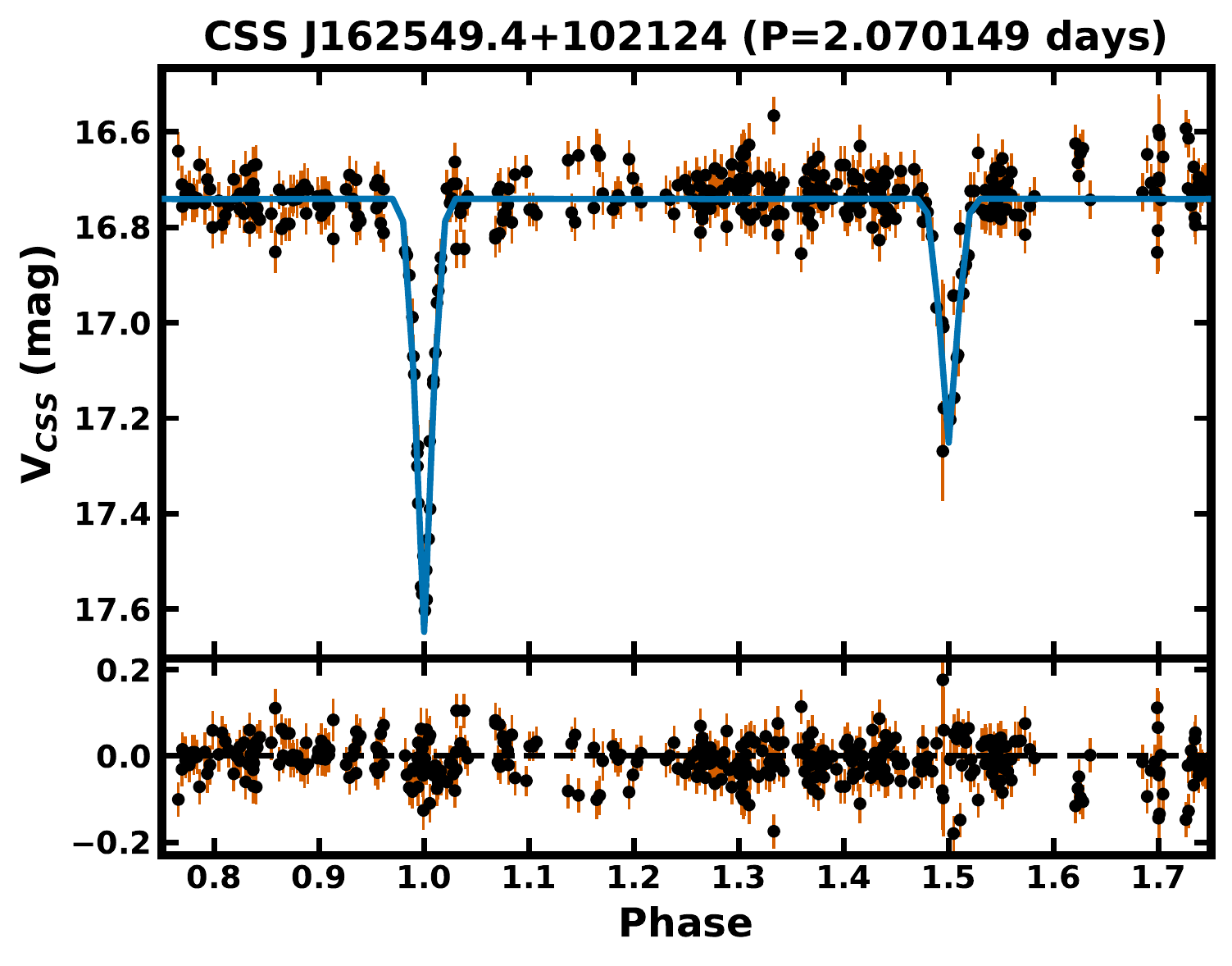}
\includegraphics[width=0.33\textwidth,height=0.3\textwidth,trim = 1mm 1mm 1mm 0mm,clip]{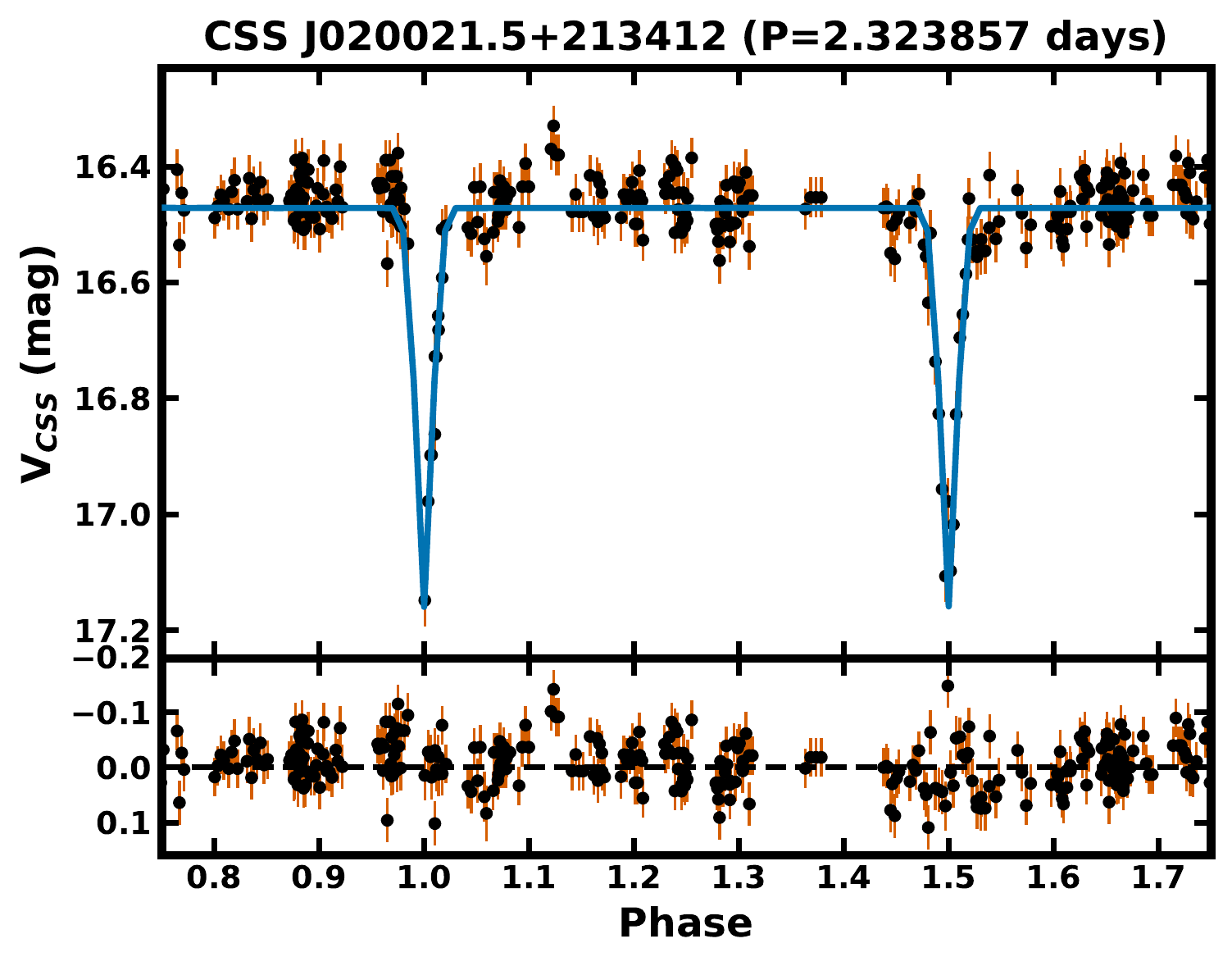}
\includegraphics[width=0.33\textwidth,height=0.3\textwidth,trim = 1mm 1mm 1mm 0mm,clip]{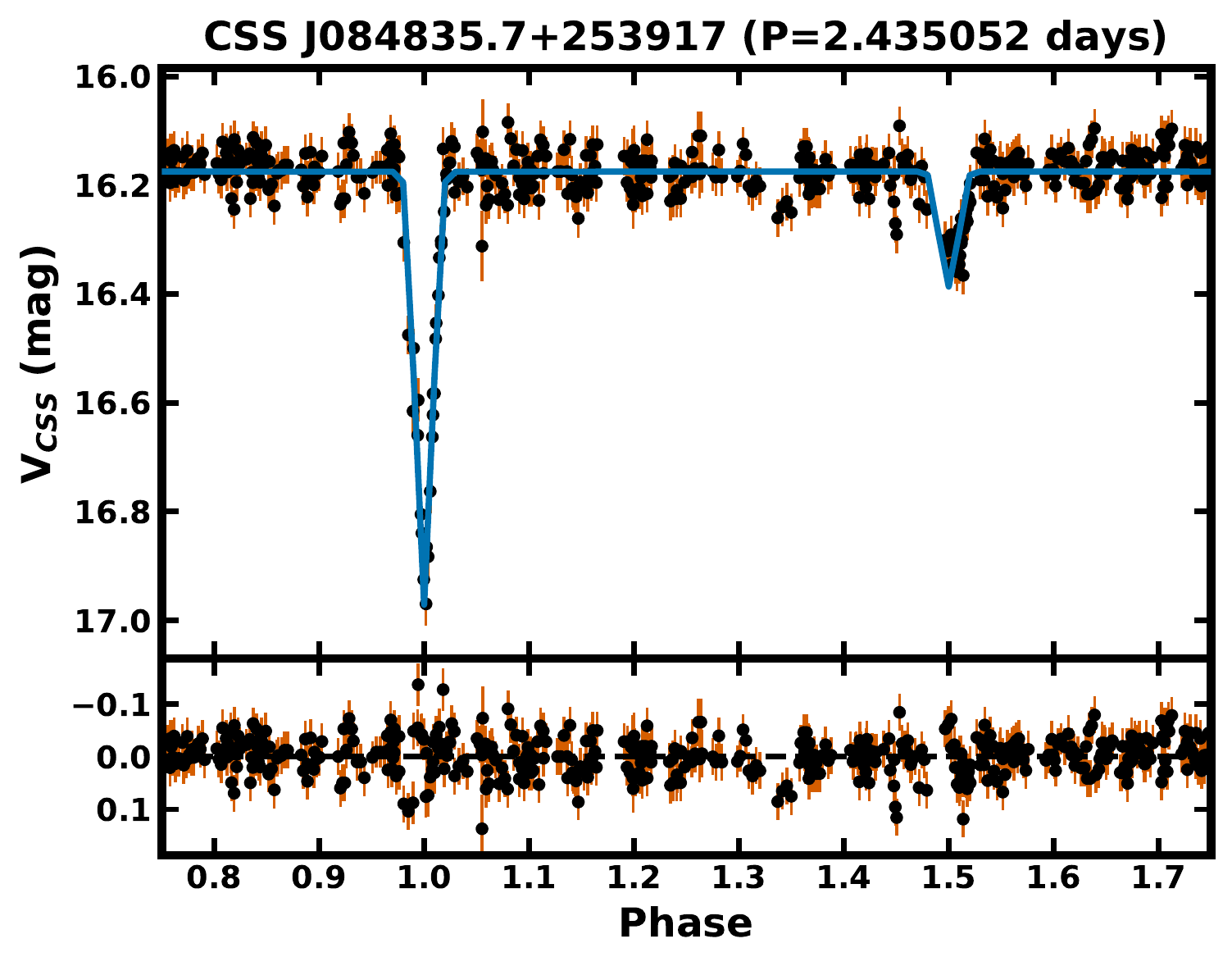}
\includegraphics[width=0.33\textwidth,height=0.3\textwidth,trim = 1mm 1mm 1mm 0mm,clip]{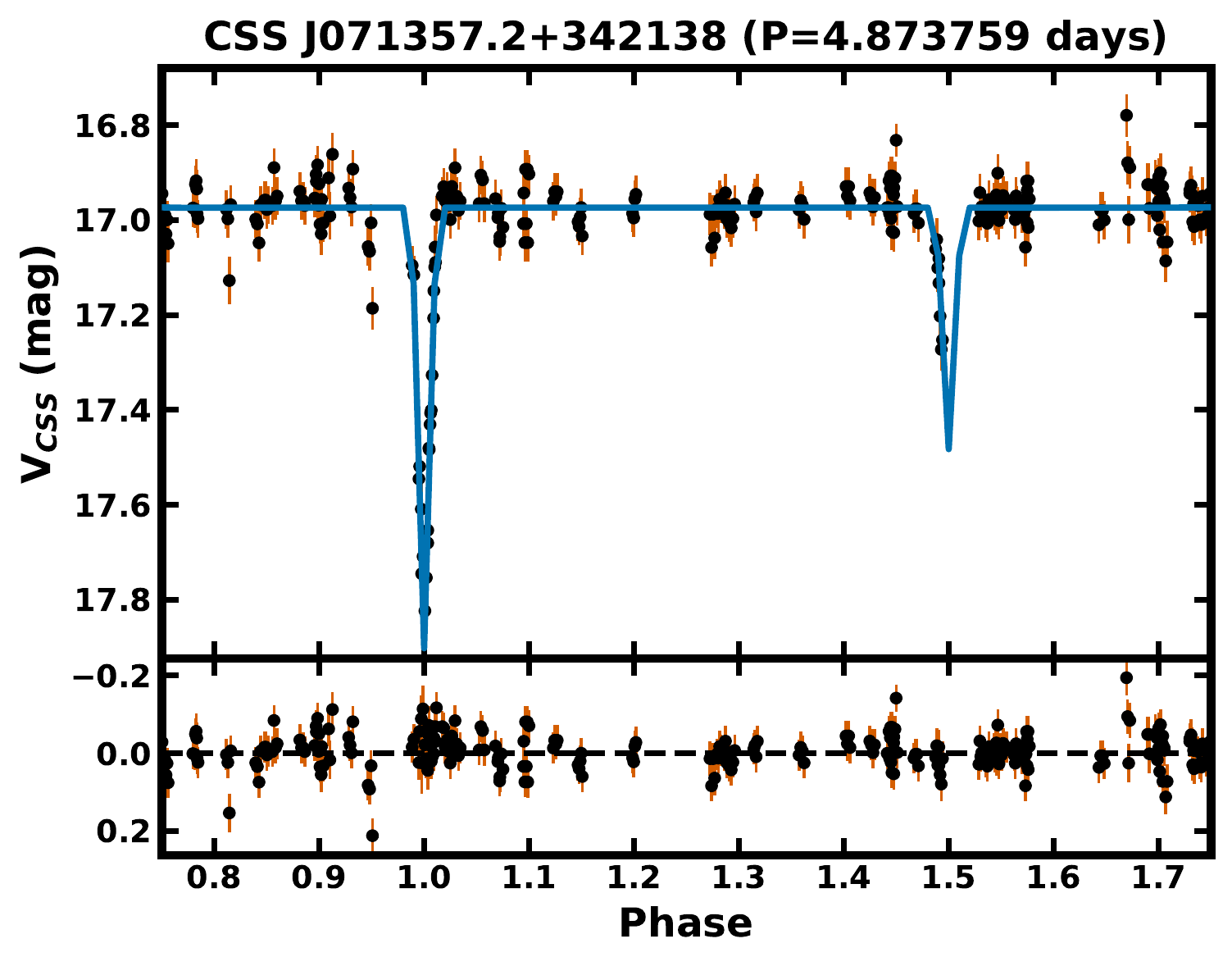}
\includegraphics[width=0.33\textwidth,height=0.3\textwidth,trim = 1mm 1mm 1mm 0mm,clip]{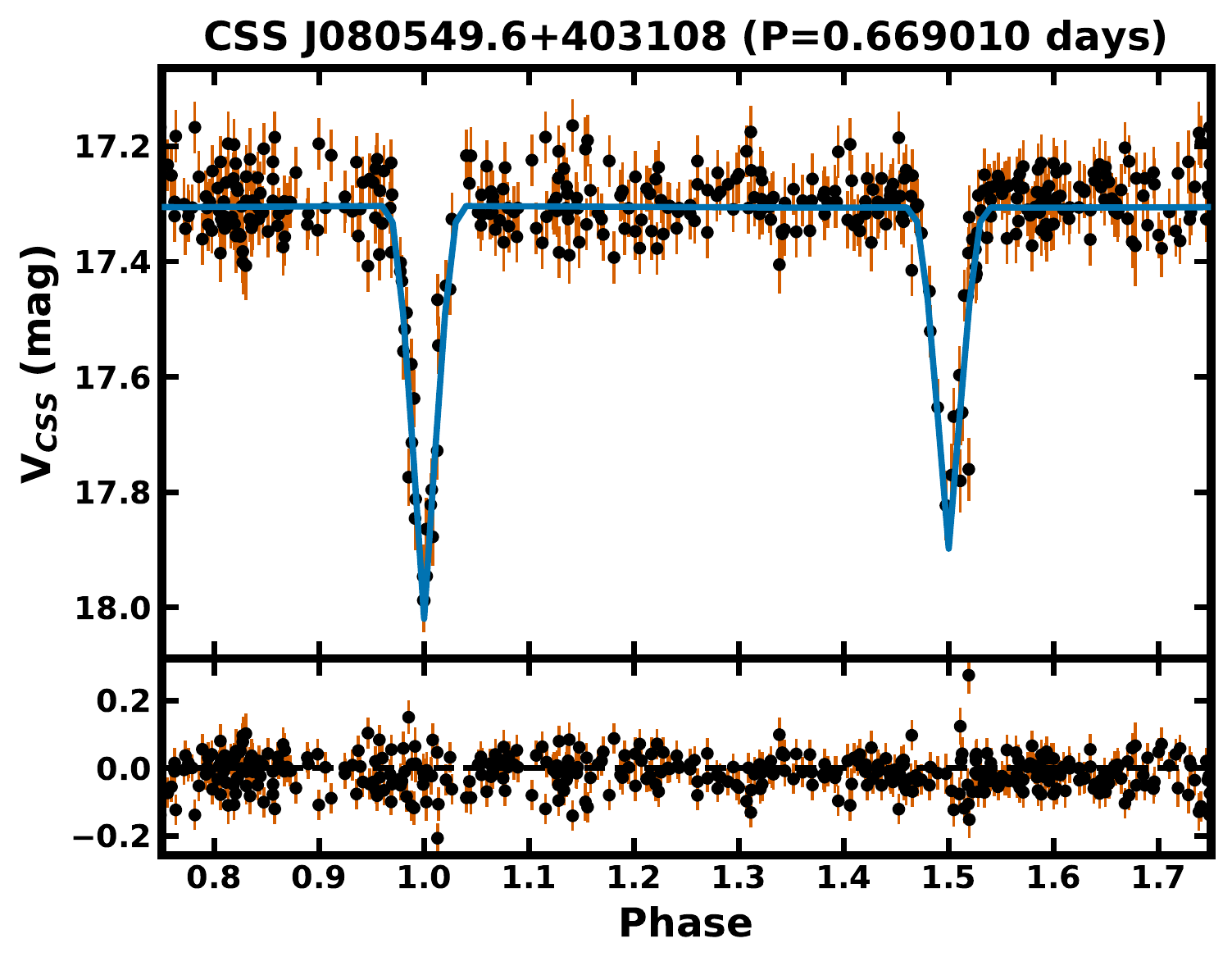}
\includegraphics[width=0.33\textwidth,height=0.3\textwidth,trim = 1mm 1mm 1mm 0mm,clip]{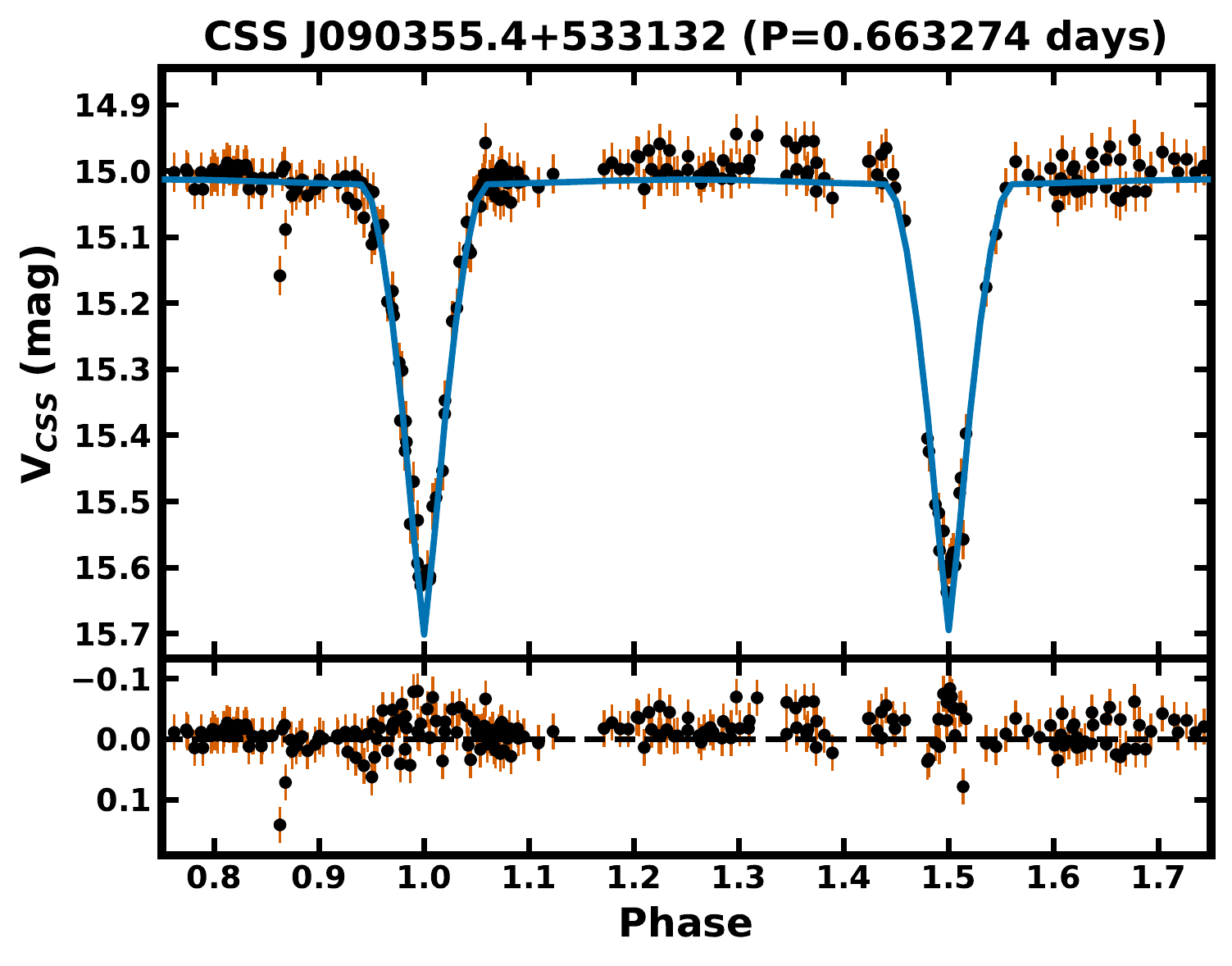}
\caption{Phase diagrams (top panels) and residuals of the fit (bottom panels) for the eight low-mass binary system candidates (for more detail see Sect. \ref{parameters}). The blue line shows the best fit solution using the parameters found in tables~\ref{tab:abspars} and \ref{tab:lmsolution_fine}.}
\label{fig:ajuste}
\end{figure*}

Having fixed the temperatures, masses, and radii of the binary components, a second WD fit was performed to refine the inclination and potential values. The results are shown in Table~\ref{tab:lmsolution_fine}. Figure ~\ref{fig:ajuste} shows the final fit obtained from Table~\ref{tab:abspars} and Table~\ref{tab:lmsolution_fine}. We verified whether the fit parameters produce a distance consistent with that provided by \citet{bailer-jones2018}. To calculate the distances, we corrected the Johnson $V$ magnitude using the extinction provided by \citet{green2018} and $R_V = 3.1$. The calculated distances (see Table~\ref{tab:lmsolution_fine}) are in agreement with those found by \citet{bailer-jones2018}. For CSS J162549.3+102124, the distance is underestimated in about $22\%$, while for the other systems the distances differ by no more than 10\% from \citet{bailer-jones2018}. These differences can be related to the $T_{eff}$ estimate since the luminosity is proportional to $T_{eff}^4$. For instance, the {\em Gaia} distance for CSS J162549.3+102124 is recovered if we assume a temperature $\sim250$ K higher than that used by us.

%%%%%%%%%%%%%%%%%%%%%%%%%%%%%%%%%%%%%%%%%%
%%%%%%%%%%%%%%%%%%%%%%%%%%%%%%%%%%%%%%%%%%

%%%%%%%%%%%%%%%%%%%%%  INFLATION
\begin{figure*}
\includegraphics[width=0.77\textwidth,height=0.57\textwidth,trim = 0mm 0mm 0mm 0mm,clip]{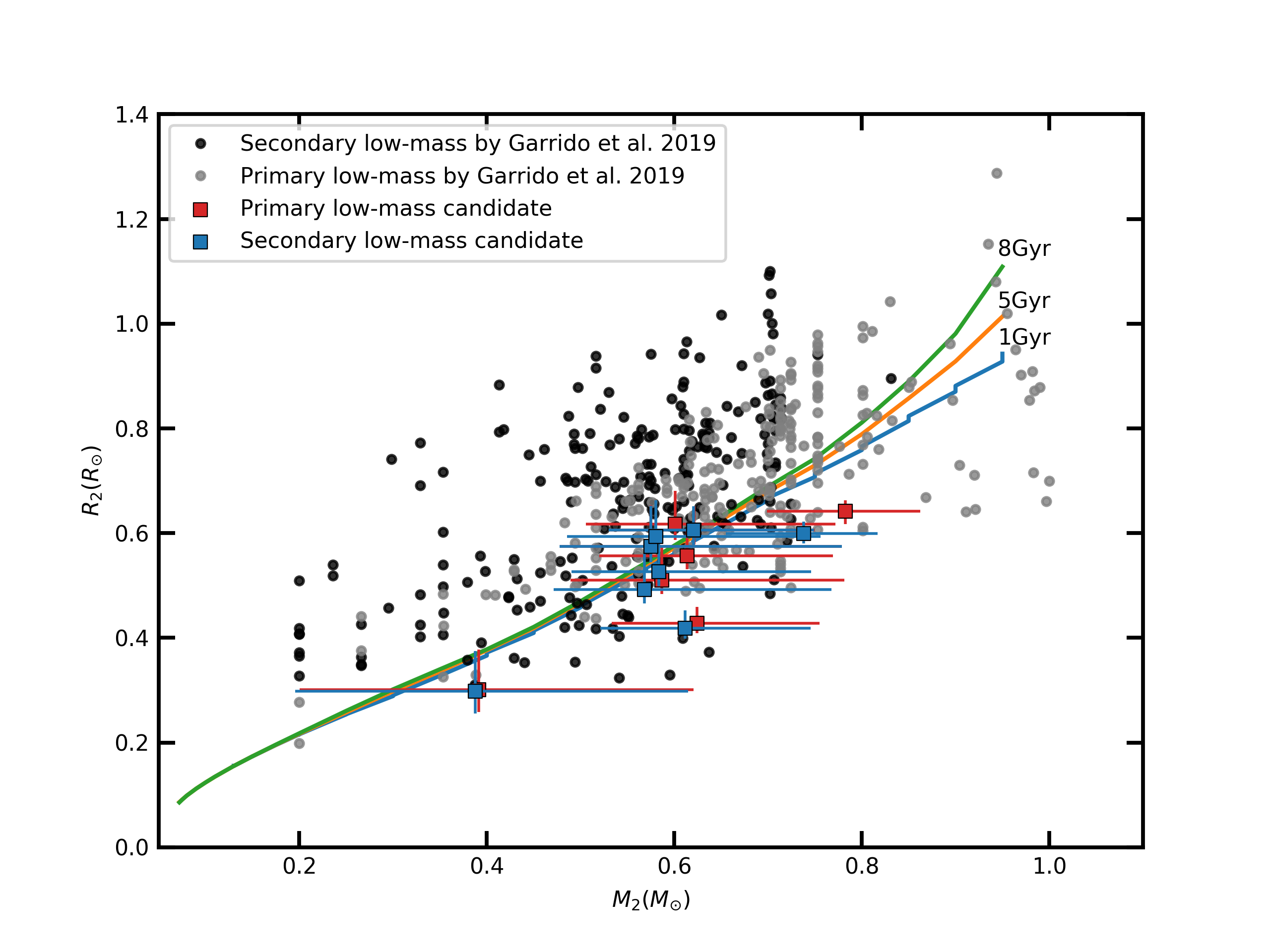}
\caption{Radius as function of mass for low-mass stars in detached binary systems. Blue and red squares represent the primary and the secondary components of our low-mass candidates, respectively. Black and gray dots represent data from \citet{garrido2019}. The theoretical models from \citet{baraffe1998} for 1, 5, and 8 Gyr are represented as blue, orange, and green solid lines, respectively.}
\label{fig:inflation}
\end{figure*}
%%%%%%%%%%%%%%%%%%%%%

\subsection{Low-mass binaries and the radius inflation problem}

\citet{garrido2019} characterized a sample of $230$ detached close-orbiting eclipsing binaries with low-mass main-sequence components ($M<1 M_\odot$), orbital periods shorter than 2 days, and temperatures below 5720~K. They suggest a trend according to which low-mass stars would have inflated radii. Besides, they found that around 61\% of the sample has the secondary star more inflated in radius than the primary.

In addition, \citet{coughlin2011} reported $95$ low-mass eclipsing binaries in the initial {\em Kepler} data release with periods as long as ten days. They presented evidence that the radius inflation of low-mass stars in binary systems decreases for longer periods: for  $P<1.0$ days, the median value of the difference between the fits and theoretical radii was about 13.0\%, whereas for $1.0<P({\rm days})<10.0$\ and $P>10.0$ days, the value was 7.5\% and 2.0\%, respectively.

Figure~\ref{fig:inflation} presents a mass-radius diagram for low-mass stars in binaries, where the objects found in this work are shown in red and blue, and the sample from \citet{garrido2019} in black and gray. The lines represent theoretical models from \citet{baraffe1998} with isochrones of 1, 5 and 8 Gyr for solar metallicity and helium abundance Y = $0.275$. 
The data from \citet{garrido2019} have most points systematically above the theoretical tracks, for both primary and secondary stars.

Our sample is well distributed around the theoretical model predictions and the median value of the difference between our results and the theoretical radii was about 4.8\%. The error bars were asymmetric, in the sense that they were larger for negative residuals (Table~\ref{tab:abspars}). Only three systems of our sample had periods shorter than two days, unlike the periods considered by \citet{garrido2019}. Even though we have a small sample, our results do not confirm the trend of inflated radii for the components of low-mass eclipsing binary systems.

\section{Conclusions}
\label{conclusions}

We reported the first determination of orbital periods for 87 EA-type eclipsing binary systems cataloged in \citet{Drake-2014} as "unknown period eclipsing binaries" (EA$_{\rm up}$), using the approach proposed by \cite{lopes2018}. This was $\sim 56$\% of the total number of objects in the (EA$_{\rm up}$) sample. We recovered periods in data sets of poorer quality compared with previous attempts.

The sample of iEA binaries selected among the EA$_{\rm up}$ objects in this work tends to show longer orbital periods, slightly brighter objects, shallower eclipses, and, of course, fewer data points than the EAs identified in the original work of \citet{Drake-2014}. The results exemplify how oversampling the natural frequency grid provided by the median sampling time can be effective in finding new objects having narrow eclipses in sparse time series. In future efforts, we intend to analyze other large surveys in an attempt to retrieve objects hidden by poor cadence and a small number of measurements.

A sub-sample of 8 eclipsing binaries with K and M spectral types was selected. We have determined the stellar parameters of the binary components by modeling their light curves with the Wilson-Devinney code. Our sample did not show clear evidence for inflated radii, either in primary or secondary components. 
As the radius inflation is expected to be stronger for systems with shorter periods, we cannot be surely assertive since there are only 3 objects with periods less than 2 days in our sample of low-mass binaries. Additional observations with radial velocities would help to constrain and refine the present individual solutions.

\section*{Acknowledgements}
% Entry for the table of contents, for this guide only
\addcontentsline{toc}{section}{Acknowledgements}
This study was financed in part by the Coordenação de Aperfeiçoamento de Pessoal de Nível Superior - Brasil (CAPES) - Finance Code 001.
C. E. F. L. acknowledges a post-doctoral fellowship from the CNPq. N. J. G. C. acknowledges support from the UK Science and Technology Facilities Council. The authors thank to MCTIC/FINEP (CT-INFRA grant 0112052700) and the Embrace Space Weather Program for the computing facilities at INPE.
C. V. Rodrigues thanks the grant \#2013/26258-4 from S\~ao Paulo Research Foundation (FAPESP) and CNPq (Proc. 303444/2018-5). Support for A. P. and M. C. is provided by Proyecto Basal AFB-170002; by the Chilean Ministry for the Economy, Development, and Tourism's Millennium Science Initiative through grant IC\,120009, awarded to the Millennium Institute of Astrophysics (MAS); and by FONDECYT project \#1171273.

%%%%%%%%%%%%%%%%%%%%%%%%%%%%%%%%%%%%%%%%%%%%%%%%%%
%%%%%%%%%%%%%%%%%%%% Data availability %%%%%%%%%%%%%%%%%%
\section*{Data availability}
The data underlying this article are available in the article and in its online supplementary material.

%%%%%%%%%%%%%%%%%%%%%%%%%%%%%%%%%%%%%%%%%%%%%%%%%%



\bibliographystyle{mnras}
\bibliography{refs} % if your bibtex file is called example.bib

%%%%%%%%%%%%%%%%%%%%%%%%%%%%%%%%%%%%%%%%%%%%%%%%%%

%%%%%%%%%%%%%%%%% APPENDICES %%%%%%%%%%%%%%%%%%%%%

\appendix
%\section{Periods and derived parameters}
%\label{sec:tables}
\section{Periods for the iEA sample}
\label{sec:tables1}

Table~\ref{tab:periodsiEA} shows the periods and the corresponding uncertainties for the iEA sample obtained in this work. These objects were found within the EA$_{\rm up}$ sample of objects with unknown periods from \citet{Drake-2014}. The methodology used is described in detail in Section~\ref{search}. The CRTS name and the periods are in the first and second columns, respectively.
$\sigma_P$ is the error in the period considering the region of the eclipses in the phase diagram. $V$(mag) is the CSS magnitude, $A$ is the amplitude and $\sigma_A$ is the amplitude error considering the region outside the eclipses in the phase diagram. $\log(T_{tot}/P)$ shows the number of cycles in the total time range of the observations in logarithmic scale. Large values of $\log(T_{tot}/P)$ represent better period estimates. For more information, see \citet{lopes2018}.

%%%%%%%%%%%%%%%%%%%%%%%
\begin{table*}
\caption{Parameters for the iEA stars identified in this work. The period $P$ in days and amplitude $A$ in magnitudes are presented with their respective errors, $\sigma_P$(d) and $\sigma_A$(mag). The $V$ is in magnitudes and $\log(T_{tot}/P)$ gives the number of cycles in logarithmic scale.}
\scriptsize
\label{tab:periodsiEA}
\begin{tabular}{lll}
\begin{tabular}{lcccccc}
\hline
Name & $P$ & $\sigma_P$ & $V$ & $A$  & $\sigma_A$ & $\log(T_{tot}/P)$\\ 
\hline
CSS J041810.3-024627 & 1.26948 & 0.00476 & 15.17 & 0.79 & 0.04 & 3.37\\
CSS J045516.1-004733 & 3.74487 & 0.07030 & 15.19 & 0.38 & 0.02 & 2.89\\
CSS J081331.4-013918 & 3.16895 & 0.01130 & 14.18 & 0.77 & 0.02 & 2.99\\
CSS J101000.7-010213 & 1.83604 & 0.01010 & 14.53 & 0.65 & 0.04 & 3.20\\
CSS J045308.6-032953 & 1.82278 & 0.10700 & 12.16 & 1.52 & 0.05 & 2.73\\
CSS J060321.6-040517 & 1.33587 & 0.00471 & 16.00 & 0.79 & 0.02 & 3.29\\
CSS J085743.8-030448 & 3.20760 & 0.02450 & 15.45 & 0.58 & 0.03 & 2.95\\
CSS J163922.3-031006 & 1.97460 & 0.00470 & 14.13 & 0.78 & 0.06 & 3.18\\
CSS J060409.8-072110 & 1.32808 & 0.00628 & 14.47 & 0.78 & 0.01 & 3.29\\
CSS J083118.6-081856 & 0.30316 & 0.00718 & 14.87 & 0.43 & 0.04 & 3.98\\
CSS J084552.4-061418 & 3.01489 & 0.00574 & 15.70 & 1.34 & 0.03 & 2.98\\
CSS J205605.9-063809 & 1.07793 & 0.00617 & 17.11 & 0.91 & 0.08 & 3.45\\
CSS J053059.3-102647 & 1.16266 & 0.00208 & 13.35 & 0.75 & 0.02 & 3.40\\
CSS J062419.5-103506 & 0.74350 & 0.00265 & 16.45 & 1.09 & 0.03 & 3.54\\
CSS J083125.1-090301 & 0.66789 & 0.00197 & 16.33 & 0.90 & 0.03 & 3.63\\
CSS J091430.0-111446 & 2.13433 & 0.01120 & 16.13 & 0.83 & 0.03 & 3.13\\
CSS J222100.4-105449 & 2.09907 & 0.00585 & 15.54 & 0.59 & 0.02 & 3.16\\
CSS J003441.8-135032 & 1.97421 & 0.00532 & 14.79 & 0.52 & 0.03 & 3.18\\
CSS J030246.9-121937 & 3.15730 & 0.00713 & 14.08 & 0.73 & 0.02 & 2.97\\
CSS J050224.2-113916 & 3.18499 & 0.00743 & 13.50 & 1.05 & 0.01 & 2.96\\
CSS J053931.4-152107 & 1.85360 & 0.02960 & 14.88 & 0.57 & 0.02 & 3.15\\
CSS J100846.9-160703 & 4.95352 & 0.06240 & 16.17 & 0.49 & 0.04 & 2.73\\
CSS J054951.2-180440 & 0.67626 & 0.01250 & 15.42 & 0.79 & 0.02 & 3.59\\
CSS J104916.3-175650 & 1.50077 & 0.02310 & 15.65 & 1.36 & 0.05 & 3.24\\
CSS J144057.5-191558 & 1.02614 & 0.00188 & 14.24 & 0.86 & 0.02 & 3.45\\
CSS J025414.6+002004 & 6.71969 & 0.16900 & 14.94 & 0.48 & 0.04 & 2.64\\
CSS J042305.8+003947 & 3.35125 & 0.03520 & 15.62 & 0.40 & 0.04 & 2.95\\
CSS J110309.4+014240 & 1.27006 & 0.01070 & 13.94 & 0.45 & 0.03 & 3.37\\
CSS J170319.1+013946 & 2.34682 & 0.03280 & 16.33 & 0.56 & 0.04 & 3.12\\
CSS J080118.6+033634 & 0.98002 & 0.00488 & 13.15 & 0.61 & 0.03 & 3.47\\
CSS J113248.7+033002 & 1.34911 & 0.03850 & 14.65 & 0.58 & 0.07 & 3.35\\
CSS J145100.7+052841 & 5.06656 & 0.09790 & 15.82 & 0.70 & 0.05 & 2.77\\
CSS J211507.1+042944 & 2.34144 & 0.02250 & 13.93 & 0.36 & 0.04 & 3.12\\
CSS J043938.6+061238 & 1.78472 & 0.01090 & 14.34 & 0.48 & 0.04 & 3.22\\
CSS J054859.3+074331 & 1.71841 & 0.00539 & 16.31 & 0.73 & 0.03 & 3.22\\
CSS J085050.6+073028 & 2.60976 & 0.00592 & 13.84 & 0.44 & 0.02 & 3.07\\
CSS J102224.7+062518 & 3.64876 & 0.01060 & 16.25 & 0.62 & 0.05 & 2.93\\
CSS J103914.0+064824 & 2.56501 & 0.04570 & 15.72 & 0.82 & 0.06 & 3.08\\
CSS J225211.1+080336 & 63.17793 & 1.65000 & 15.73 & 0.50 & 0.06 & 1.68\\
CSS J162549.3+102124 & 2.07015 & 0.04610 & 16.78 & 0.84 & 0.06 & 3.18\\
CSS J050242.7+131025 & 0.64938 & 0.01090 & 14.45 & 0.26 & 0.03 & 3.68\\
CSS J052736.9+140215 & 2.65183 & 0.00986 & 16.23 & 0.78 & 0.06 & 3.07\\
CSS J080331.2+135122 & 2.36413 & 0.02260 & 15.88 & 0.68 & 0.03 & 3.12\\
CSS J050242.7+131025 & 0.64938 & 0.01090 & 14.45 & 0.26 & 0.03 & 3.68\\
\hline
\end{tabular}

\begin{tabular}{lcccccc}
\hline
Name & $P$ & $\sigma_P$ & $V$ & $A$ & $\sigma_A$ & $\log(T_{tot}/P)$\\ 
\hline
CSS J052736.9+140215 & 2.65183 & 0.00986 & 16.23 & 0.78 & 0.06 & 3.07\\
CSS J162004.6+145346 & 3.03808 & 0.00912 & 16.21 & 0.58 & 0.03 & 3.01\\
CSS J164108.5+163433 & 1.39605 & 0.00625 & 14.15 & 0.67 & 0.06 & 3.35\\
CSS J225011.3+172418 & 2.79777 & 0.00946 & 15.37 & 0.59 & 0.04 & 3.03\\
CSS J020021.5+213412 & 2.32386 & 0.01270 & 16.49 & 0.63 & 0.04 & 3.10\\
CSS J035048.7+204955 & 2.02690 & 0.02020 & 16.60 & 0.69 & 0.04 & 3.16\\
CSS J073807.8+221414 & 5.23632 & 0.00736 & 13.56 & 0.38 & 0.01 & 2.75\\
CSS J222615.0+211301 & 2.71315 & 0.00956 & 12.88 & 1.78 & 0.10 & 3.05\\
CSS J001223.3+274350 & 1.82846 & 0.01190 & 16.04 & 0.54 & 0.04 & 3.21\\
CSS J084835.7+253917 & 2.43505 & 0.04530 & 16.20 & 0.77 & 0.05 & 3.10\\
CSS J173356.0+264846 & 2.65100 & 0.00722 & 14.12 & 0.63 & 0.02 & 3.07\\
CSS J030604.9+282408 & 1.55640 & 0.01990 & 15.80 & 0.35 & 0.04 & 3.28\\
CSS J050515.1+284725 & 1.00942 & 0.00315 & 15.57 & 0.57 & 0.03 & 3.49\\
CSS J125153.5+293917 & 3.16246 & 0.01880 & 15.36 & 0.69 & 0.02 & 2.98\\
CSS J233755.5+295554 & 2.75103 & 0.06380 & 15.40 & 0.92 & 0.05 & 3.04\\
CSS J035633.1+320912 & 3.25411 & 0.01240 & 13.34 & 0.50 & 0.02 & 2.96\\
CSS J041359.0+314056 & 3.60426 & 0.00862 & 14.96 & 0.73 & 0.03 & 2.91\\
CSS J045258.7+331809 & 2.46991 & 0.00700 & 15.78 & 1.84 & 0.03 & 3.10\\
CSS J074854.5+312748 & 6.18431 & 0.00811 & 14.47 & 0.62 & 0.03 & 2.70\\
CSS J231824.8+310818 & 2.81653 & 0.00924 & 15.72 & 1.22 & 0.04 & 3.03\\
CSS J232619.3+334509 & 2.81313 & 0.01690 & 16.16 & 0.44 & 0.04 & 3.04\\
CSS J025355.2+353950 & 3.06574 & 0.04920 & 14.17 & 0.74 & 0.03 & 2.98\\
CSS J032724.6+360153 & 1.95523 & 0.00447 & 13.75 & 0.24 & 0.03 & 3.18\\
CSS J034625.6+354612 & 1.72581 & 0.00968 & 14.20 & 0.67 & 0.05 & 3.23\\
CSS J071357.2+342138 & 4.87376 & 0.12200 & 17.01 & 0.78 & 0.06 & 2.80\\
CSS J002509.1+385544 & 2.06580 & 0.00793 & 13.58 & 0.61 & 0.01 & 3.15\\
CSS J031907.5+384354 & 1.69540 & 0.00573 & 12.95 & 1.03 & 0.05 & 3.24\\
CSS J043933.7+365854 & 2.48613 & 0.01020 & 14.28 & 0.54 & 0.03 & 3.07\\
CSS J085656.0+382028 & 2.09378 & 0.05370 & 15.57 & 0.46 & 0.04 & 3.14\\
CSS J003827.1+410334 & 3.13914 & 0.02940 & 14.41 & 0.63 & 0.02 & 2.97\\
CSS J035654.8+395231 & 2.72690 & 0.00320 & 14.08 & 1.26 & 0.01 & 3.03\\
CSS J080549.6+403108 & 0.66901 & 0.00240 & 17.33 & 0.67 & 0.06 & 3.67\\
CSS J232718.6+415044 & 2.63147 & 0.07790 & 14.68 & 1.17 & 0.06 & 3.05\\
CSS J005332.7+440226 & 1.90658 & 0.02140 & 13.88 & 1.06 & 0.03 & 3.18\\
CSS J175341.0+444623 & 1.90262 & 0.00837 & 12.71 & 0.84 & 0.05 & 3.21\\
CSS J094558.1+454814 & 0.68133 & 0.00793 & 13.61 & 0.26 & 0.03 & 3.63\\
CSS J080327.2+503948 & 1.37363 & 0.03310 & 13.75 & 0.50 & 0.02 & 3.32\\
CSS J180743.0+502014 & 0.99206 & 0.00865 & 15.81 & 1.14 & 0.11 & 3.49\\
CSS J090355.4+533131 & 0.66328 & 0.00234 & 15.10 & 0.64 & 0.03 & 3.64\\
CSS J164404.4+574227 & 0.49234 & 0.00118 & 15.09 & 0.99 & 0.03 & 3.74\\
CSS J065935.8+592538 & 1.80142 & 0.01120 & 13.14 & 0.59 & 0.03 & 3.19\\
CSS J070423.0+593108 & 0.75928 & 0.00366 & 14.59 & 0.63 & 0.02 & 3.57\\
CSS J061902.1+631324 & 0.20970 & 0.00589 & 15.75 & 0.43 & 0.03 & 4.08\\
\hline
\end{tabular}
\end{tabular}
\end{table*}

%%%%%%%%%

\section{Parameters for light curve solution}
\label{sec:tables2}
As described in Section~\ref{parameters}, the  low-mass star candidates were modeled with the WD light curve synthesis code combined with an MCMC optimization procedure. Table~\ref{tab:lmsolution} shows the parameters and the ratio of parameters found in the light curve modeling. These values were obtained as described by \citet{coughlin2011} and were used to find the absolute parameters shown in Table~\ref{tab:abspars}. The temperatures obtained by \citet{coughlin2011} were used in the WD code in a second fit, with $T_2/T_1$ fixed in order to refine $i$, $\Omega_1$ and $\Omega_2$ for a final solution (see Table~\ref{tab:lmsolution_fine}). The first and second fits made with WD code produce $T_2/T_1$  and $(R_1 + R_2)/a$ consistent to within 9\%. Using the fit parameters we can recover the distances, which are in agreement with those found in the literature. The potentials of CSS J145100.7+052843 and CSS J071357.2+342138 are relative large; this is due to the fact that there are few photometric points in the eclipse, which probably leads to underestimated values of the sum of the relative radii.

%%%%%%%%%%%%%%%%%%%%%%%%%%%%%
\begin{table*}
\scriptsize
\caption{Parameters obtained from the model fits to the light curves. The first one presents the name of the object, followed by the CSS ID, temperature of the primary $T_1$ and temperature ratio $T_1/T_2$, sum of fractional radii $(R_1 + R_2)/a$, inclination $i$ and modified Kopal potential (dimensionless) $\Omega_1$ and $\Omega_2$.}
 \label{tab:lmsolution}
 \begin{tabular}{lccccccc}
\hline   
Name & CSS ID & $T_1$ (fixed) & $T_2/T_1$ & $(R_1 + R_2)/a$ & $i^\circ$ & $\Omega_1$ & $\Omega_2$ \\
\hline
CSS J003441.8-135033 & 1012004004843 & 3896 & 0.97 $\pm$ 0.01 & 0.15 $\pm$ 0.02 & 88.08 $\pm$ 0.29 & 13.73 $\pm$ 1.45 & 14.31 $\pm$ 0.83\\
CSS J145100.7+052843 & 1104080068168 & 3954 & 0.96 $\pm$ 0.01 & 0.06 $\pm$ 0.01 & 89.67 $\pm$ 0.13 & 32.95 $\pm$ 3.77 & 31.85 $\pm$ 3.18\\
CSS J162549.4+102124 & 1109087063294 & 3828 & 0.95 $\pm$ 0.01 & 0.14 $\pm$ 0.02 & 89.62 $\pm$ 0.25 & 15.48 $\pm$ 1.10 & 15.03 $\pm$ 0.96\\
CSS J020021.5+213412 & 1121011041164 & 3819 & 1.00 $\pm$ 0.01 & 0.15 $\pm$ 0.01 & 89.58 $\pm$ 0.24 & 14.45 $\pm$ 0.62 & 14.44 $\pm$ 0.67\\
CSS J084835.7+253917 & 1126043006161 & 4510 & 0.87 $\pm$ 0.01 & 0.14 $\pm$ 0.02 & 87.81 $\pm$ 0.55 & 16.22 $\pm$ 2.34 & 14.28 $\pm$ 0.58\\
CSS J071357.2+342138 & 1135032018057 & 3868 & 0.95 $\pm$ 0.01 & 0.09 $\pm$ 0.01 & 89.77 $\pm$ 0.18 & 22.32 $\pm$ 1.70 & 21.66 $\pm$ 1.27\\
CSS J080549.6+403108 & 1140034027271 & 3512 & 0.98 $\pm$ 0.01 & 0.20 $\pm$ 0.02 & 88.94 $\pm$ 0.42 & 11.05 $\pm$ 0.75 & 10.74 $\pm$ 0.64\\
CSS J090355.4+533132 & 1152031059450 & 3962 & 1.00 $\pm$ 0.01 & 0.35 $\pm$ 0.02 & 88.72 $\pm$ 0.78 & 6.72 $\pm$ 0.45 & 6.73 $\pm$ 0.44\\
\hline  
\end{tabular}
\end{table*}
%%%%%%%%%%%%%%%%%%%%%%%%%%%%%

%%%%%%%%%%%%%%%
%\vspace{-1em}
\begin{table*}
\scriptsize
\caption{The temperatures and masses of the components were defined as described in the Section \ref{findlowmass} and the model fit to the light curve was redone in order to improve parameters such as $\Omega_1$, $\Omega_2$ and $i$. The last column is the estimated distance from the fitted parameters.}
 \label{tab:lmsolution_fine}
 \begin{tabular}{lccccccc}
\hline    
Name & CSS ID & $T_2/T_1$ (fixed) & $(R_1 + R_2)/a$ & $i^\circ$ & $\Omega_1$ & $\Omega_2$ &    $dist$ (pc) \\
\hline   
CSS J003441.8-135033 & 1012004004843 & 0.98 & 0.15 $\pm$ 0.01 & 88.29 $\pm$ 0.36 & 13.08 $\pm$ 0.94 & 14.31 $\pm$ 0.30 & 280 $\pm$ 68 \\
CSS J145100.7+052843 & 1104080068168 & 0.99 & 0.06 $\pm$ 0.01 & 89.85 $\pm$ 0.10 & 32.16 $\pm$ 3.56 & 31.70 $\pm$ 3.09 & 280 $\pm$ 57 \\
CSS J162549.4+102124 & 1109087063294 & 0.99 & 0.14 $\pm$ 0.02 & 89.88 $\pm$ 0.11 & 15.06 $\pm$ 0.74 & 14.57 $\pm$ 0.75 & 466 $\pm$ 118 \\
CSS J020021.5+213412 & 1121011041164 & 1.00 & 0.15 $\pm$ 0.02 & 89.61 $\pm$ 0.22 & 14.49 $\pm$ 0.67 & 14.51 $\pm$ 0.63 & 486 $\pm$ 124 \\
CSS J084835.7+253917 & 1126043006161 & 0.96 & 0.14 $\pm$ 0.02 & 89.04 $\pm$ 0.38 & 13.61 $\pm$ 0.33 & 15.53 $\pm$ 0.43 & 878 $\pm$ 167 \\
CSS J071357.2+342138 & 1135032018057 & 0.98 & 0.10 $\pm$ 0.01 & 89.95 $\pm$ 0.04 & 20.72 $\pm$ 0.85 & 19.99 $\pm$ 0.85 & 764 $\pm$ 192 \\
CSS J080549.6+403108 & 1140034027271 & 1.00 & 0.20 $\pm$ 0.03 & 89.19 $\pm$ 0.35 & 11.13 $\pm$ 0.53 & 10.46 $\pm$ 0.46 & 260 $\pm$ 91 \\
CSS J090355.4+533132 & 1152031059450 & 1.00 & 0.35 $\pm$ 0.03 & 88.82 $\pm$ 0.80 & 6.72 $\pm$ 0.47 & 6.73 $\pm$ 0.45 & 364 $\pm$ 78 \\
\hline  
\end{tabular}
\end{table*}
%%%%%%%%%%%%%%%

%%%%%%%%%%%%%%%%%%%%%%%%%%%%%%%%%%%%%%%%%%%%%%%%%%
%
%
% Don't change these lines
\bsp	% typesetting comment
\label{lastpage}
\end{document}